\documentclass{article}% Math and Physical Sciences Numbered Reference Style

\usepackage{graphicx}%
\usepackage{multirow}%
\usepackage{amsmath,amssymb,amsfonts}%
\usepackage{amsthm}%
\usepackage{mathrsfs}%
\usepackage[title]{appendix}%
\usepackage[dvipsnames]{xcolor}%
\usepackage{textcomp}%
\usepackage{manyfoot}%
\usepackage{booktabs}%
\usepackage{algorithm}%
\usepackage{algorithmicx}%
\usepackage{algpseudocode}%
\usepackage{listings}%
\usepackage{array}%
\usepackage[hidelinks]{hyperref}

\usepackage[section]{placeins}

\definecolor{colgreen}{RGB}{0, 103, 22}
\definecolor{colred}{RGB}{193,18,31}
\definecolor{colblue}{RGB}{0, 69, 124}

\usepackage{pgfplots}
    \pgfplotsset{
    }

\usepackage{tikz}
\usepackage{ifthen}
\usepackage{xstring}

\numberwithin{equation}{subsection}
\numberwithin{table}{subsection}

\newtheorem{definition}{Definition}

\usepackage{xspace}                 % Needed for correct space in macros
\newcommand{\toolname}{\textsc{CombOL}\xspace}

\newcommand{\countseq}[2]{
  \\\,\\ #1: \texttt{#2, \dots}\\
}

\newcommand{\seq}           {\ensuremath{\textrm{SEQ}}}
\newcommand{\mset}          {\ensuremath{\textrm{MSET}}}
\newcommand{\cyc}           {\ensuremath{\textrm{CYC}}}
\newcommand{\diag}          {\ensuremath{\mathrm{\Delta}}}
\newcommand{\atom}              {\ensuremath{\textrm{\textbf{z}}}}
\newcommand{\uatom}             {\ensuremath{\textrm{\textbf{u}}}}
\newcommand{\uatommark}         {\textcolor{colgreen}{\uatom}}
\newcommand{\vatom}             {\ensuremath{\textrm{\textbf{v}}}}
\newcommand{\vatommark}         {\textcolor{colred}{\vatom}}
\newcommand{\watom}             {\ensuremath{\textrm{\textbf{w}}}}
\newcommand{\watomark}          {\textcolor{colblue}{\watom}}

\newcommand{\class}[1]          {\ensuremath{\mathcal{#1}}}             % Notation for comb. classes
\newcommand{\emptyclass}        {\ensuremath{\textrm{\textbf{1}}}}
\newcommand{\aclass}            {\class{A}}
\newcommand{\bclass}            {\class{B}}
\newcommand{\cclass}            {\ensuremath{\mathcal{C}}}              % Generic combinatorial class

\newcommand{\btree}             {\ensuremath{\mathcal{B}}}

    \newcommand{\uptonone}  {\ensuremath{n}}
    \newcommand{\uptochiral} {\ensuremath{t}}
    \newcommand{\uptoez} {\ensuremath{e}}
    \newcommand{\uptoboth} {{\uptoez\uptochiral}}

    \newcommand{\ca}        {\ensuremath{\mathcal{A}_{\uptochiral}}}      % Alkanes, up to stereo
    \newcommand{\cat}       {\ensuremath{\mathcal{A}_{\uptonone}}}      % Alkanes, general
    \newcommand{\ce}        {\ensuremath{\mathcal{E}_{\uptoboth}}}      % Alk+olf, up to stereo
    \newcommand{\cet}       {\ensuremath{\mathcal{E}_{\uptoez}}}      % Alk+olf, up to cis-trans stereo
    \newcommand{\cec}       {\ensuremath{\mathcal{E}_{\uptochiral}}}      % Alk+olf, up to tetrahedral stereo
    \newcommand{\cetc}      {\ensuremath{\mathcal{E}_{\uptonone}}}      % Alk+olf, general

    \newcommand{\cau}        {\ensuremath{\ca^*}}      % Alkanes, counting u, no stereo
    \newcommand{\caut}       {\ensuremath{\cat^*}}    % Alkanes, counting u, stereo
    \newcommand{\ceu}        {\ensuremath{\ce^*}}    % Alkanes and alkenes, counting u, no stereo
    \newcommand{\ceut}       {\ensuremath{\cet^*}}  % -||-, counting u, chiral stereo
    \newcommand{\ceuc}       {\ensuremath{\cec^*}}  % -||-, counting u, cis-trans
    \newcommand{\ceutc}      {\ensuremath{\cetc^*}} % -||-, counting u, both stereo

    \newcommand{\dclass}             {\ensuremath{\mathcal{D}}}

    \newcommand{\dbl}               {\ensuremath{\dclass_{\uptoboth}}}
    \newcommand{\dblt}              {\ensuremath{\dclass_{\uptoez}}}
    \newcommand{\dblc}              {\ensuremath{\dclass_{\uptochiral}}}
    \newcommand{\dbltc}             {\ensuremath{\dclass_{\uptonone}}}
        \newcommand{\dblu}              {\ensuremath{\dbl^*}}
        \newcommand{\dblut}             {\ensuremath{\dblt^*}}
        \newcommand{\dbluc}             {\ensuremath{\dblc^*}}
        \newcommand{\dblutc}            {\ensuremath{\dbltc^*}}

    \newcommand{\caach}             {\ensuremath{\ca^a}}
    \newcommand{\caacht}            {\ensuremath{\cat^a}}
    \newcommand{\ceach}             {\ensuremath{\ce^a}}
    \newcommand{\ceacht}            {\ensuremath{\cet^a}}
    \newcommand{\ceachc}            {\ensuremath{\cec^a}}
    \newcommand{\ceachtc}           {\ensuremath{\cetc^a}}
    \newcommand{\dblach}            {\ensuremath{\dbl^a}}
    \newcommand{\dblacht}           {\ensuremath{\dblt^a}}
    \newcommand{\dblachc}           {\ensuremath{\dblc^a}}
    \newcommand{\dblachtc}          {\ensuremath{\dbltc^a}}

    \newcommand{\camesot}            {\ensuremath{\cat^m}}

\newcommand*{\vc}[1]{\begingroup \setbox0=\hbox{#1}\parbox{\wd0}{\box0}\endgroup}   % Vertical Center
\newcommand{\tsep}{\\[3mm]}
\newcommand{\treetablespace}{\\[1.5mm]}

\newcommand{\extract}[1]        {\ensuremath{[#1]\,}}
\newcommand{\combiso}           {\cong}                                 % Combinatorial Isomorphism

\newcommand{\oeis}[1]{\textbf{#1}}                                   % Macro for including OEIS link
\newcommand{\tz}{$ \color{gray} 0$}                                   % Zero in tables (greyed out?)

\pgfdeclarelayer{background}
\pgfsetlayers{background, main}
\newcommand{\negativecolour}{red!8}

\newcommand{\drawparent}{
	\draw(0,0) circle (0.075) [color=black, fill=black];
}

\newcommand{\drawchiral}[1]{
	\IfSubStr{#1}{chiral}{
		\node (ast) at (0.20, 0) {\textcolor{colgreen}{\large{\textbf{*}}}};
	}{}
	\IfSubStr{#1}{vchi}{
		\node (ast) at (0.20, 0) {\textcolor{colgreen}{\large{\textbf{*}}}};
	}{}
}
\newcommand{\drawfromdouble}[1]{
	\IfSubStr{#1}{fromdouble}{
		\IfSubStr{#1}{negative}{
			\path (0,0) edge [thick, double=\negativecolour] (0,0.2);
		}{
			\path (0,0) edge [thick, double] (0,0.2);
		}
	}{
		\path (0,0) edge [thick] (0,0.2);
	}
}
\newcommand{\drawnegative}[1]{
	\IfSubStr{#1}{negative}{
		\begin{pgfonlayer}{background}
			\draw [rounded corners](current bounding box.south west) rectangle (current bounding box.north east) [fill=\negativecolour, color=\negativecolour];
		\end{pgfonlayer}
	}{}
}
\newcommand{\drawsubequaltwo}[1]{   % Two nodes identical
	\IfSubStr{#1}{equal}{
		\node (eq) at (0,-1) {$=$};
	}{}
}
\newcommand{\drawsubequalthree}[1]{ % Three nodes, left two identical
	\IfSubStr{#1}{equal}{
		\node (eq) at (-0.35,-1) {$=$};
	}{}
}
\newcommand{\drawdoublesingle}[1]{      % One child connected by double
	\IfSubStr{#1}{singledouble}{
		\IfSubStr{#1}{negative}{
			\path (0,0) edge [thick, double=\negativecolour] (c1);
		}{
			\path (0,0) edge [thick, double] (c1);
		}
	}{}
}
\newcommand{\drawdoubleleft}[1]{        % Left child connected by double
	\IfSubStr{#1}{doubleleft}{
		\IfSubStr{#1}{negative}{
			\path (0,0) edge [thick, double=\negativecolour] (c1);
		}{
			\path (0,0) edge [thick, double] (c1);
		}
	}{}
}
\newcommand{\drawdoubleright}[1]{        % Left child connected by double
	\IfSubStr{#1}{doubleright}{
		\IfSubStr{#1}{negative}{
			\path (0,0) edge [thick, double=\negativecolour] (c2);
		}{
			\path (0,0) edge [thick, double] (c2);
		}
	}{}
}

\newcommand{\specbullet}[1][]{
	\vc{\tikz{
			\drawfromdouble{#1}
			\node (parent) {};
			\drawparent
		}}
}
\newcommand{\specsingle}[2][]{
	\vc{\tikz{
			\node (c1) at (0,-1) {#2};
			\path (parent.center) edge [thick] (c1);
			\drawfromdouble{#1}
			\drawdoublesingle{#1}
			\node (parent) at (0,0) {};
			\drawparent
			\drawchiral{#1}
			\drawnegative{#1}
		}}
}

\usepgflibrary{bbox}
\newcommand{\specdouble}[3][]{
	\vc{\tikz{
			\IfSubStr{#1}{noleft}{}{
				\node (c1) at (-.5,-1) {#2};
				\path (parent.center) edge [thick] (c1);
				\drawdoubleleft{#1}
			}
			\IfSubStr{#1}{noright}{}{
				\node (c2) at (.5,-1) {#3};
				\path (parent.center) edge [thick] (c2);
				\drawdoubleright{#1}
			}
			\drawfromdouble{#1}
			\node (parent) at (0,0) {};
			\drawparent
			\drawsubequaltwo{#1}
			\drawchiral{#1}
			\drawnegative{#1}
		}}
}
\newcommand{\spectriple}[4][]{
	\vc{\tikz{
			\node (c1) at (-.7, -1) {#2};
			\node (c2) at (0, -1)  {#3};
			\node (c3) at (.7, -1)  {#4};
			\drawfromdouble{#1}
			\node (parent) at (0,0) {};
			\drawparent
			\path (parent.center) edge [thick] (c1);
			\path (parent.center) edge [thick] (c2);
			\path (parent.center) edge [thick] (c3);
			\drawsubequalthree{#1}
			\drawchiral{#1}
			\drawnegative{#1}
		}}
}
\newcommand{\speccycle}[4][]{
	\vc{\tikz{
			\drawfromdouble{#1}
			\node (parent) at (0,0) {};
			\drawparent
			\draw (0,-.775) ellipse (.7 and .25) [thick, lightgray];
			\IfSubStr{#1}{negative}{
				\draw (-.7, -.85) circle (.27) [fill=\negativecolour, \negativecolour];
				\draw (0, -1) circle (.27) [fill=\negativecolour, \negativecolour];
				\draw (.7, -.85) circle (.27) [fill=\negativecolour, \negativecolour];
			}{
				\draw (-.7, -.85) circle (.27) [fill=white, white];
				\draw (0, -1) circle (.27) [fill=white, white];
				\draw (.7, -.85) circle (.27) [fill=white, white];
			}
			\node (c1) at (-.7, -.8) {#2};
			\node (c2) at (0, -1)  {#3};
			\node (c3) at (.7, -.8)  {#4};
			\path (parent.center) edge [thick] (c1);
			\path (parent.center) edge [thick] (c2);
			\path (parent.center) edge [thick] (c3);
			\drawchiral{#1}
			\drawnegative{#1}
		}}
}

\newcommand{\combdrawenv}[1]{
	\,\\
	#1
	\,\\\bigskip
}

\newcommand{\caFormula}{\begin{split}\ensuremath{
		\ca & = \atom + \atom \times \ca + \atom \times \mset_2(\ca) + \atom \times \mset_3(\ca)
		}\end{split}}
\newcommand{\caDrawing}{
	\combdrawenv{
		\begin{array}{ccccccccc}
			\ca
			 & = & \atom
			 & + & \atom \times \ca
			 & + & \atom \times \mset_2(\ca)
			 & + & \atom \times \mset_3(\ca)
			\treetablespace
			 & = & \specbullet
			 & + & \specsingle{\ca}
			 & + & \specdouble{\ca}{\ca}
			 & + & \spectriple{\ca}{\ca}{\ca}
		\end{array}
	}
}

\newcommand{\catFormula}{\begin{split}\ensuremath{
		\cat & = \atom + \atom \times \cat + \atom \times \cat \times \cat + \atom \times \cyc_3(\cat)
		}\end{split}}
\newcommand{\catDrawing}{
	\combdrawenv{
		\begin{array}{ccccccccc}
			\cat
			 & = & \atom
			 & + & \atom \times \cat
			 & + & \atom \times \cat \times \cat
			 & + & \atom \times \cyc_3(\cat)
			\treetablespace
			 & = & \specbullet
			 & + & \specsingle{\cat}
			 & + & \specdouble{\cat}{\cat}
			 & + & \speccycle{\cat}{\cat}{\cat}
		\end{array}
	}
}

\newcommand{\ceFormula}{\begin{split}\ensuremath{
		\ce  & = \atom + \atom \times \ce + \atom \times \dbl + \atom \times \mset_2(\ce) + \atom \times \ce \times \dbl + \atom \times \mset_3(\ce) \\
		\dbl & = \atom + \atom \times \ce + \atom \times \mset_2(\ce)
		}\end{split}}
\newcommand{\ceDrawing}{
	\combdrawenv{
		&\begin{array}{cccccccccccccccccccc}
			\ce
			 & = & \atom
			 & + & \atom \times \ce
			 & + & \atom \times \dbl
			 & + & \atom \times \mset_2(\ce)
			 & + & \atom \times \ce \times \dbl
			 & + & \atom \times \mset_3(\ce)
			\tsep
			 & = & \specbullet[]
			 & + & \specsingle{\ce}
			 & + & \specsingle[singledouble]{\dbl}
			 & + & \specdouble{\ce}{\ce}
			 & + & \specdouble[doubleright]{\ce}{\dbl}
			 & + & \spectriple{\ce}{\ce}{\ce}
		\end{array}
		\treetablespace
		&\begin{array}{cccccccccccccccccccc}
			\dbl
			 & = & \atom
			 & + & \atom \times \ce
			 & + & \atom \times \mset_2(\ce)         \\
			\tsep
			 & = & \specbullet[fromdouble]
			 & + & \specsingle[fromdouble]{\ce}
			 & + & \specdouble[fromdouble]{\ce}{\ce}
		\end{array}
	}
}

\newcommand{\cetFormula}{\begin{split}\ensuremath{
		\cet  & = \atom + \atom \times \cet + \atom \times \dblt + \atom \times \cet \times \cet + \atom \times \cet \times \dblt + \atom \times \cyc_3(\cet) \\
		\dblt & = \atom + \atom \times \cet + \atom \times \mset_2(\cet)
		}\end{split}}

\newcommand{\cecFormula}{\begin{split}\ensuremath{
		\cec  & = \atom + \atom \times \cec + \atom \times \dblc + \atom \times \mset_2(\cec) + \atom \times \cec \times \dblc + \atom \times \mset_3(\cec) \\
		\dblc & = \atom + \atom \times \cec + \atom \times \cec + \atom \times \cec \times \cec
		}\end{split}}

\newcommand{\cetcFormula}{\begin{split}\ensuremath{
		\cetc  & = \atom + \atom \times \cetc + \atom \times \dbltc + \atom \times \cetc \times \cetc + \atom \times \cetc \times \dbltc + \atom \times \cyc_3(\cetc)
		\\
		\dbltc & = \atom + \atom \times \cetc + \atom \times \cetc + \atom \times \cetc \times \cetc
		}\end{split}}
\newcommand{\cetcDrawing}{
	\combdrawenv{
		&\begin{array}{cccccccccccccccccccc}
			\cetc
			 & = & \atom
			 & + & \atom \times \cetc
			 & + & \atom \times \dbltc
			 & + & \atom \times \cetc \times \cetc
			 & + & \atom \times \cetc \times \dbltc
			 & + & \atom \times \cyc_3(\cetc)
			\tsep
			 & = & \specbullet[]
			 & + & \specsingle{\cetc}
			 & + & \specsingle[singledouble]{\dbltc}
			 & + & \specdouble{\cetc}{\cetc}
			 & + & \specdouble[doubleright]{\cetc}{\dbltc}
			 & + & \speccycle{\cetc}{\cetc}{\cetc}
		\end{array}
		\treetablespace
		&\begin{array}{cccccccccccccccccccc}
			\dbltc
			 & = & \atom
			 & + & \atom \times \cetc
			 & + & \atom \times \cetc
			 & + & \atom \times \cetc \times \cetc
			\tsep
			 & = & \specbullet[fromdouble]
			 & + & \specdouble[fromdouble, noleft]{-}{\cetc}
			 & + & \specdouble[fromdouble, noright]{\cetc}{-}
			 & + & \specdouble[fromdouble]{\cetc}{\cetc}
		\end{array}
	}
}

\newcommand{\cauFormula}{\begin{split}\ensuremath{
		\cau & =
		\atom + \atom \times \cau    \\
		     & + \big(
			\atom \times \uatommark \times \mset_2(\cau) \setminus
			\atom \times \uatommark \times \diag_2(\cau)
			\big)
		+ \atom \times \diag_2(\cau) \\
		     & + \big(
			\atom \times \uatommark \times \mset_3(\cau) \setminus
			\atom \times \uatommark \times \diag_2(\cau) \times \cau
			\big)
			+ \atom \times \diag_2(\cau) \times \cau
		}\end{split}}
\newcommand{\cauDrawing}{
	\begin{split}
		\combdrawenv{
		 & \begin{array}{cccccccccccccccccccc}
				   \cau
				    & = & \atom
				    & + & \atom \times \cau
				   \tsep
				    & = & \specbullet[]     & + & \specsingle{\cau}
			   \end{array}                                  \\
		 & \begin{array}{cccccccccccccccc}
				   \phantom{\cau}
				    & +         & \big(  &
				   \atom \times \uatommark \times \mset_2(\cau)
				    & \setminus &
				   \atom \times \uatommark \times \diag_2(\cau)
				    & \big)     & +      &
				   \atom \times \diag_2(\cau)
				   \tsep
				    & +         & \Bigg( &
				   \specdouble[chiral]{\cau}{\cau}
				    & \setminus &
				   \specdouble[chiral, negative, equal]{\cau}{\cau}
				    & \Bigg)    & +      &
				   \specdouble[equal]{\cau}{\cau}
			   \end{array}                              \\
		 & \begin{array}{ccccccccccccccc}
				   \phantom{\cau}
				    & +         & \big(                                                    &
				   \atom \times \uatommark \times \mset_3(\cau)
				    & \setminus & \atom \times \uatommark \times \diag_2(\cau) \times \cau
				    & \big)
				    & +         & \atom \times \diag_2(\cau) \times \cau
				   \tsep
				    & +         & \Bigg(                                                   &
				   \spectriple[chiral]{\cau}{\cau}{\cau}
				    & \setminus &
				   \spectriple[chiral, negative, equal]{\cau}{\cau}{\cau}
				    & \Bigg)    & +                                                        &
				   \spectriple[equal]{\cau}{\cau}{\cau}
			   \end{array}
		}
	\end{split}
}

\newcommand{\cautFormula}{\begin{split}\ensuremath{
		\caut & =
		\atom + \atom \times \caut    \\
		      & + \big(
			\atom \times \uatommark \times \caut \times \caut \setminus
			\atom \times \uatommark \times \diag_2(\caut)
			\big)
		+ \atom \times \diag_2(\caut) \\
		      & + \big(
			\atom \times \uatommark \times \cyc_3(\caut) \setminus
			\atom \times \uatommark \times \diag_2(\caut) \times \caut
			\big)
			+ \atom \times \diag_2(\caut) \times \caut
		}\end{split}}

\newcommand{\ceuFormula}{\begin{split}\ensuremath{
		\ceu  & =
		\atom + \atom \times \ceu + \atom \times \dblu + \atom \times \ceu \times \dblu \\
		      & + \big(
			\atom \times \uatommark \times \mset_2(\ceu) \setminus
			\atom \times \uatommark \times \diag_2(\ceu)
		\big) + \atom \times \diag_2(\ceu)                                             \\
		      & + \big(
			\atom \times \uatommark \times \mset_3(\ceu) \setminus
			\atom \times \uatommark \times \diag_2(\ceu) \times \ceu
			\big) + \atom \times \diag_2(\ceu) \times \ceu
		\\
		\dblu & =
		\atom + \atom \times \vatommark \times \ceu                                    \\
		      & + \big(
			\atom \times \vatommark \times \mset_2(\ceu) \setminus
			\atom \times \vatommark \times \diag_2(\ceu)
			\big) + \atom \times \diag_2(\ceu)
		}\end{split}}

\newcommand{\ceutFormula}{\begin{split}\ensuremath{
		\ceut  & =
		\atom + \atom \times \ceut + \atom \times \dblut + \atom \times \ce \times \dblut \\
		       & + \big(
			\atom \times \uatommark \times \ceut \times \ceut \setminus
			\atom \times \uatommark \times \diag_2(\ceut)
		\big) + \atom \times \diag_2(\ceut)                                               \\
		       & + \big(
			\atom \times \uatommark \times \cyc_3(\ceut) \setminus
			\atom \times \uatommark \times \diag_2(\ceut) \times \ceut
			\big) + \atom \times \diag_2(\ceut) \times \ceut
		\\
		\dblut & =
		\atom + \atom \times \vatommark \times \ceut                                      \\
		       & + \big(
			\atom \times \vatommark \times \mset_2(\ceut) \setminus
			\atom \times \vatommark \times \diag_2(\ceut)
			\big) + \atom \times \diag_2(\ceut)
		}\end{split}}

\newcommand{\ceucFormula}{\begin{split}\ensuremath{
		\ceuc  & =
		\atom + \atom \times \ceuc + \atom \times \dbluc + \atom \times \ce \times \dbluc   \\
		       & + \big(
			\atom \times \uatommark \times \mset_2(\ceuc) \setminus
			\atom \times \uatommark \times \diag_2(\ceuc)
		\big) + \atom \times \diag_2(\ceuc)                                                 \\
		       & + \big(
			\atom \times \uatommark \times \mset_3(\ceuc) \setminus
			\atom \times \uatommark \times \diag_2(\ceuc) \times \ceuc
			\big) + \atom \times \diag_2(\ceuc) \times \ceuc
		\\
		\dbluc & =
		\atom + \atom \times \vatommark \times \ceuc + \atom \times \vatommark \times \ceuc \\
		       & + \big(
			\atom \times \vatommark \times \ceuc \times \ceuc \setminus
			\atom \times \vatommark \times \diag_2(\ceuc)
			\big) + \atom \times \diag_2(\ceuc)
		}\end{split}}

\newcommand{\ceutcFormula}{\begin{split}\ensuremath{
		\ceutc  & =
		\atom + \atom \times \ceutc + \atom \times \dblutc + \atom \times \ceutc \times \dblutc \\
		        & + \big(
			\atom \times \uatommark \times \ceutc \times \ceutc \setminus
			\atom \times \uatommark \times \diag_2(\ceutc)
		\big) + \atom \times \diag_2(\ceutc)                                                    \\
		        & + \big(
			\atom \times \uatommark \times \cyc_3(\ceutc) \setminus
			\atom \times \uatommark \times \diag_2(\ceutc) \times \ceutc
			\big) + \atom \times \diag_2(\ceutc) \times \ceutc
		\\
		\dblutc & =
		\atom + \atom \times \vatommark \times \ceutc + \atom \times \vatommark \times \ceutc   \\
		        & + \big(
			\atom \times \vatommark \times \ceutc \times \ceutc \setminus
			\atom \times \vatommark \times \diag_2(\ceutc)
			\big) + \atom \times \diag_2(\ceutc)
		}\end{split}}

\newcommand{\caachFormula}{\begin{split}\ensuremath{
		\caach & = \atom + \atom \times \caach + \atom \times \diag_2(\ca) + \atom \times \caach \times \diag_2(\ca)
		\\
			\caFormula
		}\end{split}}

\newcommand{\caachtFormula}{\begin{split}\ensuremath{
		\caacht & = \atom + \atom \times \caacht + \atom \times \diag_2(\cat) + \atom \times \caacht \times \diag_2(\cat)
		\\
			\catFormula
		}\end{split}}

\newcommand{\ceachFormula}{\begin{split}\ensuremath{
		\ceach  & = \atom + \atom \times \ceach + \atom \times \dblach + \atom \times \ceach \times \dblach + \atom \times \diag_2(\ce) + \atom \times \ceach \times \diag_2(\ce)
		\\
		\dblach & = \atom + \atom \times \ceach + \atom \times \mset_2(\ceach)
		\\
			\ceFormula
		}\end{split}}

\newcommand{\ceachtFormula}{\begin{split}\ensuremath{
		\ceacht  & = \atom + \atom \times \ceacht + \atom \times \dblacht + \atom \times \ceacht \times \dblacht + \atom \times \diag_2(\cet) + \atom \times \ceacht \times \diag_2(\cet)
		\\
		\dblacht & = \atom + \atom \times \ceacht + \atom \times \mset_2(\ceacht)
		\\
			\cetFormula
		}\end{split}}

\newcommand{\ceachcFormula}{\begin{split}\ensuremath{
		\ceachc  & = \atom + \atom \times \ceachc + \atom \times \dblachc + \atom \times \ceachc \times \dblachc + \atom \times \diag_2(\cec) + \atom \times \ceachc \times \diag_2(\cec)
		\\
		\dblachc & = \atom + \atom \times \ceachc + \atom \times \ceachc + \atom \times \ceachc \times \ceachc
		\\
			\cecFormula
		}\end{split}}

\newcommand{\ceachtcFormula}{\begin{split}\ensuremath{
		\ceachtc  & = \atom + \atom \times \ceachtc + \atom \times \dblachtc + \atom \times \ceachtc \times \dblachtc + \atom \times \diag_2(\cetc) + \atom \times \ceachtc \times \diag_2(\cetc)
		\\
		\dblachtc & = \atom + \atom \times \ceachtc + \atom \times \ceachtc + \atom \times \ceachtc \times \ceachtc
		\\
			\cetcFormula
		}\end{split}}

\pgfplotsset{
    PlotAxisStyle/.style={
        name=boundary,
        axis lines*=left,     % Asterisk swithces off lines
        ymin=0, ymax=0.25,
        ylabel={\%},
        xtick={0,5,10,15,20},
        minor x tick num=4,         % Minor ticks in between numbers
        ytick={0,0.05,0.1,0.15,0.2, 0.25},
        yticklabel style={
            /pgf/number format/fixed
        },
        scaled y ticks=false,
        ymajorgrids=true,
        grid style=dashed,
        legend pos=south east,
        legend cell align={left},
        legend style={
            draw=none
        },
    }
}

\pgfplotsset{
    PlotStyle/.style={
        mark=*
    }
}

\newcommand{\legendtable}[1]{
    \node[
        draw,
        fill=white,
        color=white, text=black,
        inner sep=0pt,
        above left=0.5em
    ] at (boundary.south east) {
        \small
        \begin{tabular}{clr}
            #1
        \end{tabular}
    };
}

\usepackage{authblk}

\begin{document}

\title{Enumerating Chemical Structures, Stereoisomers, and Stereogenic Units with Analytic Combinatorics}

\author[1]{Casper Asbjørn Eriksen}
\author[2,3,1]{Daniel Merkle}

\affil[1]{
    Department of Mathematics and Computer Science, %
    University of Southern Denmark
}
\affil[2]{
	Algorithmic Cheminformatics Group, Faculty of Technology %
	Bielefeld University
}
\affil[3]{
	Centre for Biotechnology (CeBiTec), %
	Bielefeld University
}

\maketitle

\section*{Abstract}
	The enumeration of chemical structures traditionally relies on graph- and group-theoretic techniques, including Pólya theory, in order to account for stereochemical symmetries.
	These methods require bespoke derivations and methods for each molecular family of interest, and quickly become very intricate for nontrivial classes.
	In this work, we demonstrate that analytic combinatorics provides a unifying, highly extensible, and relatively simple framework for enumeration problems.
	Using the symbolic method, we show that classical counting problems, such as the enumeration of acyclic alkanes and other hydrocarbon families, can be naturally solved via concise combinatorial specifications of the class of interest.
	We first demonstrate the simplicity of this approach by replicating existing results in this framework, requiring significantly less work than previously necessary, and then extend these ideas to more complex problems, counting broader classes of molecules accounting for both tetrahedral and E/Z stereoisomerism, counting the distribution of stereogenic units, and enumerating achiral and meso compounds.
	All results are made reproducible through CombOL, our open-source software package which automates the translation of combinatorial specifications into generating functions using the symbolic method, and performs both univariate and multivariate enumeration.

\section{Introduction} \label{sec:introduction}
The enumeration of chemical structures - such as alkanes, polyenoids, benzenoids, polyhex hydrocarbons, or fullerenes - has been an active area of research in both chemistry and mathematics for more than a century.
As early as Cayley, and continually through the work of Henze, Blair, Pólya, Fujita, and others, chemists and mathematicians have sought methods for counting structurally distinct molecules and their stereoisomers.
These efforts have produced a diverse collection of combinatorial techniques, ranging from direct graph-theoretic approaches and recurrence relations to symmetry-correcting methods based on group actions and Pólya theory.

Acyclic hydrocarbons in particular have served as a classical subject of study in enumeration.
Their tree-like nature permits elegant mathematical formulations, but the incorporation of stereochemical considerations rapidly complicates traditional counting approaches.
One influential family of methods is George Pólya's counting approach, originally introduced in the 1930s to systematically account for symmetries by combining group actions and generating functions \cite{Polya_1937_german}.
More recently, S. Fujita \cite{Fujita_book} has published extensively on mathematical stereochemistry, producing a single-author book and numerous papers, using combinatorial and group-theoretic ideas inspired by or built on Pólya theory.

With this work, we aim to introduce analytic combinatorics as an alternative methodological foundation to the field of enumerative chemistry.
Analytic combinatorics provides a framework within which classes of structures, such as classes of molecules, can be specified and automatically translated into associated generating functions.

This approach not only simplifies many classical counting problems, but also extends naturally to broader classes of chemical structures and multivariate generating functions without requiring fundamentally different techniques.
Analytic combinatorics, though building on earlier foundations in enumerative combinatorics, was systematically developed and firmly established as a distinct field by Flajolet and Sedgewick in 2009 \cite{purple_book}, integrating classical methods on generating functions with advanced tools from complex analysis to provide both enumerative and asymptotic insights into classes of \emph{combinatorial structures}.
Although generating functions are central to both Pólya theory-based methods and analytic combinatorics, the latter offers a mathematical framework consisting of symbolic, set-theoretical constructions via the \emph{symbolic method} and systematic asymptotic expansions that can simplify and extend previous counting results, especially for large-scale or multivariate problems, such as the multivariate enumeration of stereocentres and other stereogenic units.

In this paper, we revisit many of the same classes that Cayley, Pólya, Fujita, and others have enumerated by classical group-theoretical approaches: alkanes and their stereoisomers.
This demonstrates how the symbolic method of analytic combinatorics readily reproduces known counting sequences.
We then extend this method to handle more complex molecular families (e.g., combining tetrahedral stereogenic centres and stereogenic double bonds, tracking E/Z stereoisomerism, enumerating stereogenic units, and enumerating meso molecules) by the same approach.

All examples presented here can be reproduced using our open-source software package, \toolname, which automates many of the tasks involved: parsing a combinatorial specification, performing the symbolic method, deriving the associated generating functions, and performing single- and multivariate enumerations.
Although this work is concerned with chemical enumerations, \toolname is designed for general combinatorial classes and can be applied to enumeration problems in any domain.

\section{Methods} \label{sec:methods}
In this section, we outline the methodological framework used throughout the paper.
This framework is based on analytic combinatorics and allows us to encode various classes of molecules via combinatorial specifications, which can then automatically be translated into generating functions via the symbolic transfer theorem.
We begin with a brief introduction to analytic combinatorics and its core constructions, followed by a description of our implementation in the CombOL software library.

\subsection{Analytic Combinatorics} \label{sec:ac}
In this section, we will give a brief and informal introduction to the field of analytic combinatorics.
The basic objects of study are \emph{combinatorial structures} and \emph{combinatorial classes} thereof, as well as the \emph{specifications} which formally describe these classes.
This section will introduce these concepts and, superficially, their mathematical inner workings.
However, an understanding of the mathematical background is not required to apply analytic combinatorics to enumeration problems, and \toolname can be used with nothing more than a \emph{specification} for the class in question as input.
For a more thorough introduction to the field, we recommend the foundational book \emph{Analytic Combinatorics} by P. Flajolet and R. Sedgewick~\cite{purple_book}.
In this work, we use the notation for combinatorial classes introduced by Flajolet and Sedgewick, but many sources use the notation of Bergeron et al.~\cite{Bergeron_book}.
Pivoteau et al. \cite{newton} also provide a good introduction to the field, bridging both notations.

Analytic Combinatorics supports a variety of applications beyond exact enumeration.
Additionally, the framework can be used for uniform random sampling of combinatorial classes (Boltzmann sampling)~\cite{boltzmann} and for asymptotic analysis of the classes via singularity analysis of the generating functions~\cite{purple_book}.
In the present work, however, we restrict ourselves to enumeration.
Our focus is to demonstrate the accessibility of automated workflows enabled by analytic combinatorics, while asymptotic analysis in general requires substantial manual effort.

% Basic defs
The most basic objects of concern are classes of combinatorial (finite) structures.
\begin{definition}[Combinatorial class] \label{def:class}
	A combinatorial class, or class, $\cclass$, is a countable set of finite structures, along with a size function on the structures $|\cdot| : \cclass \to \mathbb{N}$, such that the class contains a finite number of structures of any given size.
\end{definition}

\begin{figure}[b]
	\centering
	\includegraphics[width=.8\textwidth]{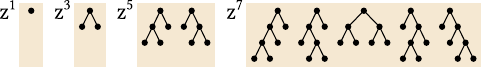}
	\caption{
    Some structures in the combinatorial class $\class{B}$ of binary trees, counting the number of vertices.
    The corresponding generating function is
    $\displaystyle \class{B}(z) = z + z^3 + 2z^5 + 5z^7 + \cdots,$
    which encodes the counting sequence
    $\{ b_n \} = 0, 1, 0, 1, 0, 2, 0, 5, \cdots.$
	}
	\label{fig:gf_btree_example}
\end{figure}

Common examples of classes of combinatorial structures and their corresponding size functions include strings (length), binary trees (number of nodes, Fig.~\ref{fig:gf_btree_example}), graphs (number of vertices), and molecules (number of atoms).

Given a class $\cclass$ and a structure $c \in \cclass$, the size of the structure is $|c|$.
We denote by $\cclass_n$ the subset of structures of $\cclass$ of size $n$, $\cclass_n = \{ c \in \cclass \mid |c| = n \}$.
The corresponding lowercase letter denotes the number of structures of each of these classes, $c_n = |\cclass_n|$.
The sequence $\{c_n\}$ then represents the number of structures of each size, and is called the \emph{counting sequence} of the class $\cclass$.
This is encoded by the \emph{generating function} of the class.

\begin{definition}[Generating function]
	\label{def:gf}
	Given a class $\cclass$, the corresponding generating function, $C(z)$, is the formal power series encoding the counting sequence $\{c_n\}$:
	$$  C(z)
		= \sum_{c \in \cclass} z^{|c|}
		= \sum_{n=0}^{\infty} c_n z^n
		= c_0 + c_1 z + c_2 z^2 + c_3 z^3 + \cdots
		,
	$$
	where each coefficient $c_n$ gives the number of structures of size $n$, and $z \neq 0$ is a formal variable.
\end{definition}

The extraction operator, $\extract{z^n} C(z)$, is used to denote the $n$th coefficient of the generating function, where
$$
	\extract{z^n} C(z)
	= c_n
	.
$$
We say that combinatorial classes $\aclass$ and $\bclass$ are \emph{combinatorially isomorphic}, denoted $ \aclass \combiso \bclass$, iff they have identical counting sequences.
This concept is important for modelling approaches, in which a class of structures is modelled in terms of symbolic constructions with the aim of obtaining a class which is combinatorially isomorphic to the original.
Enumerating the latter class then yields the enumeration of the former.
Fig. \ref{fig:mol_to_tree} provides an example of the combinatorial isomorphism exploited in this work.

% Finite classes
The simplest classes are those with a finite number of structures.
These are called \emph{finite} or \emph{primitive} classes, and they are necessary as a basic atom from which to specify complex classes.
The finite classes commonly used are the \emph{neutral class} $\emptyclass$, which consists of a single structure with size $0$, and the \emph{atomic class} $\atom$, which consists of a single structure of size $1$.
Their respective generating functions follow directly from Definition~\ref{def:gf}: $\emptyclass(z) = 1$, $\atom(z) = z$.

% Recursive classes
More complex classes can be constructed from these atoms.
Constructions are recursively defined by their \emph{specification}, a formula based on combinatorial constructions.
Each construction is simply a set-theoretical operation, describing the construction of a class from other classes, and the corresponding \emph{symbolic transfer theorem} gives the generating function of the constructed class based on the generating functions of the constituent classes.
This method is known as the \emph{symbolic method}, and allows us to obtain the generating function, and therefore the counting sequence, of any constructible class.

A classical example is the class of ordered binary trees, $\btree$\footnote{
	Here, the class of binary trees is the class of full, rooted, plane binary trees.
}%
, which may be described recursively:
\emph{``a binary tree is either a node, or a node with two binary trees attached''}.
This description can be translated to a combinatorial specification:
$$
	\btree = \atom + (\atom \times \btree \times \btree)
	,
$$
where the `$+$' and `$\times$' operators are defined in the usual set-theoretic sense as the disjoint union and Cartesian product, informally corresponding to the concepts ``\emph{or}'' and ``\emph{and}'', respectively.
The structures of this class, up to size seven, and the generating function, are presented in Fig.~\ref{fig:gf_btree_example}.
We will proceed to describe the most common combinatorial constructions, based on the descriptions given in \cite{purple_book}.

% Union
\noindent
\textbf{Disjoint union}\quad
The class $\class{A} + \class{B}$ is defined as the disjoint union (combinatorial sum) of structures from classes $\class{A}$ and $\class{B}$.
In case $\class{A}$ and $\class{B}$ intersect, we may construct the disjoint union by appending to the structures of either class distinct zero-cost structures, $1_\class{A}$ and $1_\class{B}$, such that
$$
	\class{A} + \class{B}
	= \left( \class{A} \times \{1_\class{A}\} \right)
	\cup \left( \class{B} \times \{1_\class{B}\} \right)
	.
$$

% Product
\noindent
\textbf{Product}\quad
The class $\class{A} \times \class{B}$ is the Cartesian product containing ordered tuples of the form $(a,b), \, a \in \class{A}, \, b \in \class{B}$.
We denote the $n$-fold product of a class $\class{A}$ with itself by $\class{A}^n$.

% Difference
\noindent
\textbf{Difference}\quad
The difference construction $\aclass \setminus \bclass$ requires that $\bclass \subseteq \aclass$, and constructs the set of structures which are in $\aclass$ but not in $\bclass$.
While this operator is rarely explicitly described in existing literature on analytic combinatorics, it is a natural extension, and proves useful for enumerating multivariate classes.

% Sequence
\noindent
\textbf{Sequence}\quad
The sequence construction, denoted $\seq(\aclass)$ corresponds to the infinite union
$\emptyclass + \aclass + \aclass^2 + \aclass^3 + \cdots$,
i.e. the class of all (possibly empty) sequences consisting of structures from $\aclass$.
Note that if the class $\aclass$ contains structures of size $0$, $\seq(\aclass)$ is an invalid construction; this class would contain an infinite number of structures of size $0$, which contradicts Definition~\ref{def:class}.
We call invalid specifications like these \emph{inadmissible}, and restrict ourselves to \emph{admissible} specifications for the remainder of this article.

% Cycle
\noindent
\textbf{Cycle}\quad
The cycle construction, $\cyc(\aclass)$, is similar to the sequence construction, but only constructs one sequence among the set of sequences which are equivalent up to cyclic permutation.
In the present application, this construction will prove useful in representing the symmetries of tetrahedral stereogenic centres.

% Mset
\noindent
\textbf{Multiset}\quad
The multiset construction, denoted $\mset(\aclass)$, constructs the class of all multisets (sets allowing repetition) with structures drawn from $\aclass$.
As with the cycle construction, the multiset construction can be considered a modification of the sequence construction, being equivalent to the sequence up to arbitrary permutation.

% Bounded constructions
\noindent
\textbf{Bounded constructions}\quad
The sequence, cycle, and multiset constructions permit tuples of arbitrary size.
However, we may wish to describe more limited classes.
We denote by $\seq_k(\aclass)$, $\cyc_k(\aclass)$, $\mset_k(\aclass)$ the constructor producing sequences (resp. cycle, multiset) of exactly $k$ structures.

% Diagonal
\noindent
\textbf{Diagonal}\quad
In some cases we want to specifically generate a number of identical structures.
For a tuple consisting of $k$ identical structures from $\aclass$, we write $\diag_k(\aclass)$, where:
$$
	\diag_k(\aclass) = \{  (\alpha, \cdots, \alpha) \mid \alpha \in \aclass \}
	.
$$
The class $\diag_k(\aclass)$ is called the diagonal, as it contains only the `diagonal' entries of the $k$-fold Cartesian product of \aclass.

% Symbolic transfer
\subsubsection{The Symbolic Method}
\label{sec:symbolicmethod}
The \emph{symbolic method} is the translation of a combinatorial specification, described using the constructions mentioned above, directly into a generating function.
Any combinatorial class with an admissible symbolic specification admits a generating function which can be used to analyze the class.
The generating function can be derived by recursively applying the \emph{symbolic transfer theorems}.
These give, for each construction, the corresponding generating function in terms of the generating functions of the constituent classes.
For example, the generating function of the union of classes $\aclass$ and $\bclass$, $\aclass + \bclass$ is simply $A(z) + B(z)$, and for the product $\aclass \times \bclass$, the generating function is $A(z)\,B(z)$.
Table~\ref{tab:constructions} gives the symbolic transfer theorems of each construction.

Returning to our previous example of the class of binary trees, $\bclass = \atom + (\atom \times \bclass \times \bclass)$, we can apply the symbolic transfer theorems of the product and union constructions:
\begin{align*}
	B(z) & = z + (zB(z)^2)
	\\
	     & = \frac{1 - \sqrt{(1-4z^2)}}{2z}
	\intertext{Performing a series expansion on this expression yields the expected formal power series with the Catalan numbers (OEIS \oeis{A000108}) as coefficients:}
	     & = z + z^3 + 2z^5 + 5z^7 + 14z^9 + 42z^{11} + \cdots
	.
\end{align*}

In the case of binary trees, which is a relatively simple class, we were able to easily derive the closed-form generating function.
It is not always possible to solve these equations algebraically (e.g. quinary trees due to Abel's impossibility theorem), in which case numerical methods may be used to evaluate the coefficients up to an arbitrary degree.

\toolname, the software package used in this paper, is able to apply the symbolic transfer theorems and enumerate classes automatically, so explicit knowledge of the symbolic transfer theorems is not required.

\begin{table}[h]
	\centering
	\begin{tabular}{lll}
		 & \textbf{Notation}
		 & \textbf{Symbolic transfer thm.}
		\\
		\hline
		\multicolumn{3}{l}{\textbf{Finite classes}}                                                              \\
		Neutral class
		 & $\emptyclass$
		 & $1$
		\\
		Atom
		 & $\atom$
		 & $z$
		\\
		Atoms
		 & $\uatom$, $\vatom$, ...
		 & $u$, $v$
		\\
		\hline
		\multicolumn{3}{l}{\textbf{Basic Constructions}}
		\\
		Union
		 & $\class{A} + \class{B}$
		 & $A(z) + B(z)$
		\\
		Product
		 & $\class{A} \times \class{B}$
		 & $A(z) \cdot B(z)$
		\\
		Difference
		 & $\aclass \setminus \bclass$
		 & $A(z) - B(z)$
		\\
		Sequence
		 & $\seq(\class{A})$
		 & $\left(1-A(z)\right)^{-1}$
		\\
		Cycle
		 & $\cyc(\class{A})$
		 & $\sum_{k=1}^\infty \left(\phi(k)/k\right) \log \left( 1 - A(z^k)\right)^{-1}  $
		\\
		Multiset
		 & $\mset(\class{A})$
		 & $\exp \left( \sum_{k=1}^{\infty} 1 / k \, A(z^k) \right)$
		\\
		\hline
		\multicolumn{3}{l}{\textbf{$k$-constructions}}
		\\
		$k$-sequence
		 & $\seq_k(\aclass)$
		 & $A(z)^k$
		\\
		$k$-cycle
		 & $\cyc_k(\aclass)$
		 & $ [u^k] \sum_{l=1}^{\infty} \frac{\phi(l)}{l} \log \frac{1}{1- u^l A(z^l)} $
		\\
		$k$-multiset
		 & $\mset_k(\aclass)$
		 & $ [u^k] \exp \left( \frac{u}{1} A(z) + \frac{u^2}{2} A(z^2) + \frac{u^3}{3} A(z^3) + \cdots \right) $
		\\
		Diagonal
		 & $\diag_k(\aclass)$
		 & $A(z^k)$
	\end{tabular}
	\caption{
		Overview of the basic combinatorial constructions and their corresponding generating function.
		$\phi(k)$ denotes the Euler totient function, the number of integers $i < k$ for which $i$ and $k$ are relatively prime.
	}
	\label{tab:constructions}
\end{table}

\subsubsection{Generating Functions in Multiple Variables}
\label{sec:mgf}
\begin{figure}[h]
	\centering
	\includegraphics{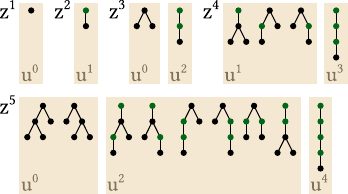}
	\label{fig:ubtrees}
	\caption{
		The unary-binary trees, as defined by Eq. \ref{eq:ubtree_u}, up to size $5$ in $z$.
		Unary nodes (those counted with $u$) are marked in green, and structures are grouped by their size in $z$ and $u$.
	}
\end{figure}

In the previous descriptions of generating functions, we have described them as formal power series in one variable, $z$, which corresponds to the \emph{size} parameter.
This approach may be generalized to an arbitrary number of variables, which count particular features of the structures.
The size of a structure is then defined as a tuple of natural numbers, each corresponding to the size in a particular variable.
We will restrict this introduction to two variables; in this setting, the second variable is commonly called the \emph{cost}.
We then refer to the size of a structure $a$ as $|a|_z$, and its cost as $|a|_u$.
Consider the class of unary-binary trees, trees in which each internal node can have one \emph{or} two children, which is described by the specification:
\begin{equation} \label{eq:ubtree}
	\class{UB} = \atom + (\atom \times \class{UB}) + (\atom \times \class{UB} \times \class{UB})
	.
\end{equation}
Say we are interested in the distribution of the number of nodes with a single child.
We introduce a second variable, $u$, such that generating functions have the form $\aclass(z,u)$, and an atom $\uatom$ with $\uatom(z,u) = u$.
We modify the specification by adding a $u$-atom to the construction producing nodes with a single child:
\begin{equation} \label{eq:ubtree_u}
	\class{UB} = \atom + (\atom \times \uatom \times \class{UB}) + (\atom \times \class{UB} \times \class{UB})
	.
\end{equation}
Applying the symbolic method to produce a generating function, and expanding this generating function in terms of $z$ yields the formal power series:
$$ \class{UB}(z,u) =
	z +  u z^2 + (1 + u^2) z^3 + (3u + u^3) z^4 + (2 + 6u^2 + u^4) z^5 + \cdots .
$$
In this representation, each coefficient $\extract{z^n}UB(z,u)$ is itself a polynomial in $u$ which describes, among the structures of size $n$, the distribution of structures according to their size in $u$, the number of nodes with a single child.
Alternatively, the relationship between $z$ and $u$ might be represented as a matrix:
\begin{table}[h!]
	\centering
	\begin{tabular}{l|rrrrr}
		\,             & $\mathbf{1}$ & $\mathbf{u}$ & $\mathbf{u^2}$ & $\mathbf{u^3}$ & $\mathbf{u^4}$
		\\ \hline
		$\mathbf{z}$   & 1            & \tz          & \tz            & \tz            & \tz            \\
		$\mathbf{z^2}$ & \tz          & 1            & \tz            & \tz            & \tz            \\
		$\mathbf{z^3}$ & 1            & \tz          & 1              & \tz            & \tz            \\
		$\mathbf{z^4}$ & \tz          & 3            & \tz            & 1              & \tz            \\
		$\mathbf{z^5}$ & 2            & \tz          & 6              & \tz            & 1              \\
	\end{tabular}
\end{table}
The above series expansion is called a \emph{horizontal expansion} in reference to the matrix representation, in which the coefficients of the `outer' series are represented by matrix rows.
Alternatively, a \emph{vertical expansion} may be performed by expanding on $u$.
An application of this could be computing the counting sequence of unary-binary trees with exactly one node with a single child by the vertical expansion $ \extract{u^1}UB(z,u) $, which corresponds to taking the `$u$' column in the matrix.
The trees counted in this table are illustrated in Fig. \ref{fig:ubtrees}.

\subsection{Implementation}    \label{sec:combol}
The methods presented in the previous section lend themselves to be performed automatically.
Given a symbolic mathematics suite and numerical algorithms, procedures can be implemented to automatically perform enumeration and analysis of a wide range of combinatorial classes in a generic fashion.
This observation is the basis for the \emph{Combinatorial Objects Library} (\toolname), a Rust-based library with a Python interface for specifying, analyzing, and enumerating combinatorial classes~\cite{combol}.
\toolname relies on the Symbolica library for symbolic computation.
For specifications with generating functions which cannot be solved in closed form, \toolname uses an iterative method based on~\cite{newton} to numerically compute the coefficients.
\toolname is open-source and available at \href{https://gitlab.com/casbjorn/combol}{gitlab.com/casbjorn/combol}.
A Python notebook providing the code to reproduce the results replicated in this paper using \toolname is available at \href{https://gitlab.com/casbjorn/molenum}{gitlab.com/casbjorn/molenum}.

The initial terms of the univariate molecule and stereoisomer enumerations were validated independently up to size $z = 12$ by a chemical graph-based pipeline: for each size, the graph grammar framework of \cite{mod} was used to exhaustively generate all constitutional isomers (without stereochemical information).
For each such structure, Marvin (ChemAxon) was used to compute all admissible stereoisomers, including both tetrahedral and E/Z stereoisomers where applicable, and the resulting counts agree with the corresponding coefficients produced by \toolname.
Beyond $z = 12$, exhaustive validation with the graph grammar pipeline becomes computationally infeasible.

\section{Enumeration of Classes of Molecules}
\label{sec:enum}
In this section, we will describe our approach to enumerating various classes of molecules according to their carbon content using the \emph{symbolic method} and \toolname.
In particular, we are interested in classes of molecules which have a tree-like structure, as these lend themselves to be intuitively described using combinatorial constructions.

\subsection{Enumerating Hydrocarbons}
\label{sec:enum_hydrocarbons}
\begin{figure}[H]
	\centering
	\includegraphics[scale=1.25]{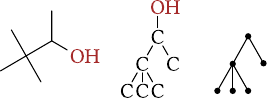}
	\caption{
		A demonstration of the combinatorial isomorphism between monosubstituted acyclic hydrocarbons and rooted trees.
		Left: The structural formula of 3,3-Dimethyl-2-butanol.
		Centre: An alternative graphical arrangement of the structure, illustrating the tree-like nature of the molecule.
		Right: The rooted tree corresponding to the structure.
		Combinatorially, the structure may be described as:
		$$
			(\atom, \cyc((\atom, \cyc(\atom, \atom, \atom)), \atom))
			.
		$$
	}
	\label{fig:mol_to_tree}
\end{figure}

We consider the class of \emph{monosubstituted acyclic hydrocarbons}.
Hydrocarbons consist only of hydrogen and carbon atoms, which allow us to model only the carbon atoms, leaving free valences filled by hydrogen atoms only implicitly represented.
Restricting ourselves to \emph{acyclic} hydrocarbons permits modelling the molecules as trees.
Finally, \emph{monosubstituted} hydrocarbons have a single functional group attached; this functional group serves as the `root' of the tree in this model.
Note that we also do not explicitly represent this functional group, which only serves to `fix' the root of the tree.
Alternatively, these enumerations also apply to acyclic hydrocarbon \emph{ligands}, with the root of the tree representing the connection to the central metal atom of the complex.
While it would also be a valid approach to explicitly represent the root, this results in a more complex specification, and the two options are \emph{combinatorially isomorphic}, so we proceed with the simplest model.
An example of this combinatorial model of a molecule is provided in Fig.~\ref{fig:mol_to_tree}.
In keeping with the mathematical terminology of trees, we use the terminology of `parents', `siblings', and `children' to describe the relationships between atoms.

We will begin by describing the simplest cases, replicating existing results.
In later sections we will expand our models to more complex classes, obtaining novel results.

\subsubsection{Alkanes}
We begin by enumerating the class of alkanes, fully saturated hydrocarbons \emph{up to tetrahedral stereoisomerism}, i.e. without any stereochemical considerations, treating stereoisomers as identical structures.
The model is therefore that of \emph{unordered} trees, that is, trees are equivalent up to the ordering of subtrees.
With this in mind, our class of alkanes up to tetrahedral stereoisomerism can be described as ``unordered trees, in which each node has up to three children''.
This description then translates to a specification for the following combinatorial class, denoted by $\ca$, here presented with explanatory diagrams:
\begin{equation}
	\caDrawing
	\label{eq:ca}
\end{equation}
In the notation of combinatorial classes used in this paper, the subscript denotes the types of symmetry with respect to which the structures are identified.
In this case, we take the class \emph{up to} tetrahedral stereoisomerism (\uptochiral), that is, molecules which differ by tetrahedral stereoisomerism are counted as one.
The above construction reads as: ``a monosubstituted acyclic alkane is either: a node; a node with an alkane attached; a node with two alkanes attached; or a node with three alkanes attached''.
Note that the $\mset$ construction is used, in which the order of the selected sub-structures does not matter.
Using \toolname, we can conveniently enumerate the class up to, e.g. twenty terms, producing the following sequence: (further terms are given in Appendix \ref{sec:appendix_enum}).
\countseq{\ca}{
	0, 1, 1, 2, 4, 8, 17, 39, 89, 211, 507, 1238, 3057, 7639, 19241, 48865, 124906, 321198, 830219, 2156010, 5622109}

This sequence is recorded in the On-Line Encyclopedia of Integer Sequences (OEIS)~\cite{oeis} as \oeis{A000598}.
The earliest enumeration of this class in 1874 is due to A. Cayley~\cite{Cayley_1874}, who enumerated six terms (although only five terms correctly) by systematic counting.
The sequence was further expanded by C. M. Blair and H. R. Henze in 1931~\cite{Henze_Blair_1931} using a system of recurrence relations.
Further methods for enumerating this class include S. Fujita~\cite[Ch. 8.3]{Fujita_book} using a method based on Pólya theory.

Using \toolname, we automatically computed the terms, giving only the specification as input.
Note that our use of analytic combinatorics in enumerating this class is not novel; in fact, this class is used as an example by P. Flajolet and R. Sedgewick~\cite[Ch. 7]{purple_book}.
What we aim to demonstrate is that this simple method can be translated to enumerate much more complex classes without adding too much complexity to the specification.

\begin{figure}[h!]
	\centering
	\includegraphics{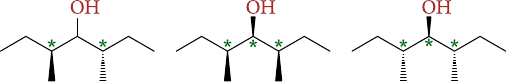}
	\caption{
		Three stereoisomers of \emph{3,5-dimethylheptan-4-ol} with tetrahedral stereogenic centres marked by asterisks, demonstrating the complexity of correctly accounting for stereoisomerism.
		Whether the carbon at position 4 (the root) is stereogenic depends on the configuration of its subtrees, specifically, carbons 3 and 5.
		In the latter two cases, the carbon in position 4 is a \emph{pseudoasymmetric centre}, as two of its ligands are enantiomorphic~\cite{nomenclature}.
	}
	\label{fig:chiral_depends_on_subtree}
\end{figure}

% Alkanes
A more fine-grained enumeration of the class can be performed by taking stereoisomerism into account, discriminating stereoisomers from each other.
This is achieved by replacing the $\mset$ construction from the previous example (which ignores the order), with order-variant alternatives.
First, recall that a carbon atom is a tetrahedral stereogenic centre iff it has four different ligands, in which case it can exist in exactly two configurations.
In our model, for any carbon atom, the `parent' substituent is necessarily different from the others, as it includes the unique functional group.
In the case of a carbon with two subtrees, the third child substituent is an implicit hydrogen atom, and the composition of the two subtrees therefore matters: if they are different, the given carbon atom is a tetrahedral stereogenic centre.
The corresponding term in the specification is then defined by the ($\times$) construction: $(\atom \times \cat \times \cat)$, rather than the $(\atom \times \mset_2(\cat))$ construction in the previous class.
Similarly, a carbon with three subtrees is a tetrahedral stereogenic centre if they are all distinct.
The cycle construction ($\atom \times \cyc_3(\cat)$) captures this notion by counting only structures up to cyclic permutations.
Note that whether the subtrees are different also depends recursively on the configuration of any tetrahedral stereogenic centre within the subtree, as demonstrated in Fig.~ \ref{fig:chiral_depends_on_subtree} - this is a subtle detail which adds a lot of complexity to conventional counting methods, but in our case, this is implicitly handled by the combinatorial constructions.

These considerations bring us to the following specification:
\begin{equation}
	\catDrawing
	\label{eq:cat}
\end{equation}
With the subscript indicating that we are enumerating this class, taken up to no symmetries (\uptonone).
Enumeration of $\cat$ yields the following sequence:
\countseq{\cat}{
	0, 1, 1, 2, 5, 11, 28, 74, 199, 551, 1553, 4436, 12832, 37496, 110500, 328092, 980491, 2946889, 8901891, 27012286, 82300275}

\begin{figure}[h]
	\centering
	\includegraphics{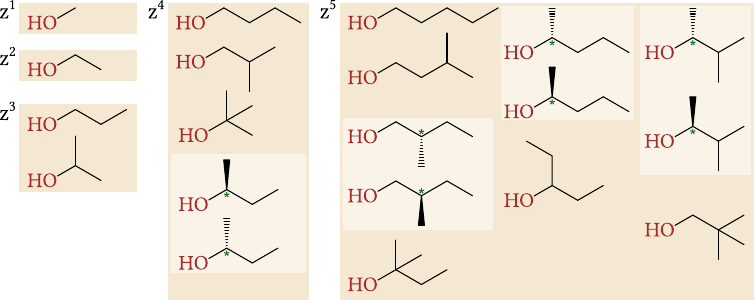}
	\caption{
		The class of monosubstituted acyclic alkanes, up to size 5.
		Tetrahedral stereogenic centres are marked with asterisks.
		Sets of structures which are equivalent up to tetrahedral stereoisomerism are grouped together.
		Counting these sets as one gives the enumeration of Eq. \ref{eq:ca}, while counting them separately gives the enumeration of Eq. \ref{eq:cat}.
	}
	\label{fig:mol_enum_a}
\end{figure}

This sequence is also recorded in the OEIS as \oeis{A000625}.
This sequence was first enumerated in a 1932 paper by C. M. Blair and H. R. Henze by means of an intricate series of recurrence relations~\cite{Blair_Henze_1932}.
Another method for describing this class was derived by G. Pólya in 1937~\cite{Polya_1937_german}\footnote{Translated from German as \cite{Polya_Read_1987}.}.
Fig.~\ref{fig:mol_enum_a} shows the concrete structures of the enumeration of alkanes, up to size 5.

\subsubsection{Alkanes and Olefins}
We now move beyond established results by considering a more complex class, including monosubstituted olefins, hydrocarbons with at least one double bond (IUPAC recommends the term \emph{olefins} for hydrocarbons with one or more double bonds \cite{gold_book}), and excluding \emph{allenes}, hydrocarbons with two consecutive double bonds due to their relative instability.

We start by taking the class of alkanes and olefins, up to stereoisomerism - now both \emph{tetrahedral} (\uptochiral) and \emph{E/Z} (\uptoez) stereoisomerism, denoting this class $\ce$.
Incorporating double bonds requires distinguishing carbon atoms connected to their parent by a single bond from those connected by a double bond, as the latter has only two valence electrons left.
We can define multiple classes which are mutually specified in terms of each other using a \emph{system of combinatorial classes}, such as the following, in which carbon atoms connected to their parent atom by a single bond are denoted by $\ce$, and those connected by a double bond as $\dbl$:
\begin{equation}
	\begin{split}
		\ceDrawing
		\label{eq:ce}
	\end{split}
\end{equation}
Here, the class $\ce$ can have up to three subtrees, unless one of them is connected by a double bond, in which case only two are permitted, while $\dbl$ has only two free valence electrons, and may have at most two subtrees.
When enumerating, we choose to enumerate the class $\ce$, as the `root' atom is assumed to connect to the functional group by a single bond.
This results in the sequence:
\countseq{\ce}{
	0, 1, 2, 5, 14, 41, 127, 410, 1353, 4576, 15732, 54874, 193639, 690206, 2480944, 8983629, 32738382, 119980438, 441907696, 1634900899, 6072852892}

The description and enumeration of this class is a novel result of this work.
The combinatorial specification of this class is only slightly more complex than the previous results, yet the mathematical framework of analytic combinatorics allows us to easily generate results for classes such as this, which would require a substantial amount of work to describe using more traditional methods such as Pólya theory.

As with alkanes, we can modify the previous specification to distinctly count stereoisomers, denoting this class $\cetc$.
In addition to tetrahedral stereoisomerism, we also need to account for E/Z stereoisomerism at double bonds.
This requires duplicating the $(\atom \times \cetc)$ term of $\dbltc$: when a carbon is connected to its parent by a double bond, whether the subtree $\cetc$ is placed in the E or Z configuration needs to be counted separately.
Even as these semantically different constructions have identical definitions, they are connected by the set-theoretic disjoint union, and are therefore counted separately.
\begin{equation}
	\begin{split}
		\cetcDrawing
		\label{eq:cetc}
	\end{split}
\end{equation}
Enumerating on $\cetc$, we obtain the following counting sequence:
\countseq{\cetc}{
	0, 1, 2, 6, 20, 69, 257, 988, 3900, 15738, 64581, 268672, 1130693, 4804808, 20587994, 88854104, 385898644, 1685306216, 7396489878, 32605260910, 144302156477}

Here, we note that the counts for this class are equal to or greater than the counts not considering stereoisomerism, as expected.
The two enumerations diverge at size 3, representing the smallest stereochemical molecule in these classes.

\subsection{Counting Stereogenic Units}
\label{sec:count_stereo}
In the previous constructions, we have taken stereogenic units into account in order to avoid overcounting the number of structures.
We now extend our approach to explicitly count the distribution of stereogenic units at various sizes using multivariate generating functions.
% Asymmetric centres terminology
For precision, we first explicitly define the terminology used in this paper.
A \emph{stereogenic unit} refers to any grouping within a molecule that gives rise to stereoisomerism;
a \emph{tetrahedral stereogenic centre} is a tetrahedral carbon atom bonded to four distinguishable ligands such that interchanging two ligands produces a stereoisomer;
a \emph{stereogenic double bond} is a $C=C$ double bond for which each atom bears two distinguishable ligands, permitting two configurations.
Note that tetrahedral stereogenic centres include \emph{pseudoasymmetric centres} - these have exactly two enantiomorphic ligands, and are stereogenic centres, but not strictly chirality centres.
When relevant, we identify pseudoasymmetric centres separately.

As before, we denote by $\atom$ the variable indicating the size of a structure, the number of carbons;
by $\uatom$ the variable counting the number of tetrahedral stereogenic centres;
and by $\vatom$ the variable counting the number of stereogenic double bonds.
We can then construct a combinatorial specification such that the coefficients of the generating function in $\atom$ are themselves polynomials in $\uatom$ and/or $\vatom$; the coefficients then describe the number of structures of size $\atom$ with $\uatom$ tetrahedral stereogenic centres and $\vatom$ stereogenic double bonds.

\subsubsection{Alkanes}
We start with the simplest case, considering the alkanes up to stereoisomerism as in Eq. \ref{eq:ca}.
In this case, the only stereogenic units are tetrahedral stereogenic centres, so a bivariate specification in $\atom$ and $\uatom$ is used.
We will first consider the atoms with two children, corresponding to the term $\atom \times \mset_2(\ca)$ of Eq. \ref{eq:ca}.
Changing this to $\atom \times \uatom \times \mset_2(\ca)$ is incorrect, as the case where both instances of $\ca$ in the multiset are identical will lead to the atom being incorrectly counted as a tetrahedral stereogenic centre.
This overcounting can be corrected by explicitly accounting for this case using the `diagonal' construction presented in sec. \ref{sec:ac}, where $\diag_n(\aclass)$ represents $n$ identical structures chosen from $\aclass$.
Consider the following modified construction:
\begin{equation*}
	\big(
	\atom \times \uatom \times \mset_2(\ca)
	\setminus
	\atom \times \uatom \times \diag_2(\ca)
	\big) + \atom \times \diag_2(\ca)
	.
\end{equation*}
Here, we explicitly exclude the problematic case from being counted with $\uatom$, and instead add it back, this time counting only with $\atom$.
Similarly for the case of an atom with three children:
\begin{equation*}
	\big(
	\atom \times \uatom \times \mset_3(\ca)
	\setminus
	\atom \times \uatom \times \diag_2(\ca) \times \ca
	\big) + \atom \times \diag_2(\ca) \times \ca
	.
\end{equation*}
Note that we do not explicitly represent the case with three identical subtrees - this is a subset of the cases with two identical subtrees.
We now arrive at the full combinatorial specification, describing alkanes, not considering stereochemistry, and counting the tetrahedral stereogenic centres with $\uatom$:
\begin{align}
	\cauDrawing
	\label{eq:cau}
\end{align}
In the drawing accompanying the specification, the tetrahedral stereogenic centres marked with $\uatom$ are denoted with an asterisk, and excluded/subtracted terms are displayed with a red background.
The diagonal construction $\diag$ is illustrated by the equals sign.
\\

As this class depends on two variables, the result is a formal power series in both $\atom$ and $\uatom$.
Expanding on either variable yields a formal power series with coefficients that are formal power series in the other. In this case, we get (expanding on $z$):
\begin{equation*}
	\cau(z,u) = z + z^2 + 2z^3 + (3 + u)z^4 + (5 + 3u)z^5 + (8 + 8u + u^2)z^6 + \dots
	.
\end{equation*}
Here, the coefficient of $z^n$ describes the class ${\cau}_n$, and the distribution of structures within this class based on their size in $u$, the number of tetrahedral stereogenic centres.
Since any given structure has its size in $u$ upper bounded by its size in $z$ (a hydrocarbon cannot have more tetrahedral stereogenic centres than it does carbon atoms), when we expand on $z$, the coefficients of the formal power series are finite polynomials in $u$.
Note that setting $u = 1$ in the generating function above yields the same generating function as Eq. \ref{eq:ca} by design,
$$
	\left. \cau(z,u) \right|_{u=1} = \ca(z)
	.
$$
This fact is useful as a `sanity check': whether we count tetrahedral stereogenic centres should not make a difference in the total number of structures of a given size;  the class $\cau$ with the $\uatom$-atoms removed is combinatorially isomorphic to $\ca$.
We can concisely represent a prefix of the expanded generating function in tabular form, as in Table \ref{tbl:cau}.
\begin{table}[h]
	\footnotesize
	\begin{tabular}{l|rrrrr|r}
		\,                & $\mathbf1$ & $\mathbf{u}^{1}$ & $\mathbf{u}^{2}$ & $\mathbf{u}^{3}$ & $\mathbf{u}^{4}$ & $\Sigma$ \\ \hline
		$\mathbf{{z}}$    & 1          & \tz              & \tz              & \tz              & \tz              & 1        \\
		$\mathbf{z}^{2}$  & 1          & \tz              & \tz              & \tz              & \tz              & 1        \\
		$\mathbf{z}^{3}$  & 2          & \tz              & \tz              & \tz              & \tz              & 2        \\
		$\mathbf{z}^{4}$  & 3          & 1                & \tz              & \tz              & \tz              & 4        \\
		$\mathbf{z}^{5}$  & 5          & 3                & \tz              & \tz              & \tz              & 8        \\
		$\mathbf{z}^{6}$  & 8          & 8                & 1                & \tz              & \tz              & 17       \\
		$\mathbf{z}^{7}$  & 14         & 20               & 5                & \tz              & \tz              & 39       \\
		$\mathbf{z}^{8}$  & 23         & 46               & 19               & 1                & \tz              & 89       \\
		$\mathbf{z}^{9}$  & 39         & 102              & 63               & 7                & \tz              & 211      \\
		$\mathbf{z}^{10}$ & 65         & 220              & 184              & 37               & 1                & 507      \\
		$\mathbf{z}^{11}$ & 110        & 461              & 503              & 154              & 10               & 1238     \\
	\end{tabular}
	\label{tbl:cau}
	\caption{
		The enumeration of the class $\cau$, in which each row enumerates the structures of a given size, and each column the structures with a given number of tetrahedral stereogenic centres.
		The column denoted $\Sigma$ gives the sum of the row; this enumeration is equivalent to that of $\ca$.
	}
\end{table}
Expanded tables are given in Appendix~\ref{sec:appendix_enum}.
In this table, the column labeled \textbf{1} is the number of structures with size $0$ in $\uatom$ - those with no tetrahedral stereogenic centres (note that this is not the same count as that of \emph{achiral} molecules, which will be discussed in Section~\ref{sec:achiral}).
We observe that the first chiral structures start appearing at size $4$ - these are the R and S configurations of \emph{2-butanol}, the smallest chiral alcohol.

We similarly extend the specification of alkanes presented in Eq. \ref{eq:cat} to count tetrahedral stereogenic centres (this time presented without the explanatory figures for brevity).
\begin{align}
	\cautFormula
	\label{eq:caut}
\end{align}
This enumeration is given in Table \ref{tbl:caut}.
\begin{table}[h]
	\footnotesize
	\begin{tabular}{l|rrrrr|r}
		\,                & $\mathbf1$ & $\mathbf{u}^{1}$ & $\mathbf{u}^{2}$ & $\mathbf{u}^{3}$ & $\mathbf{u}^{4}$ & $\Sigma$ \\ \hline
		$\mathbf{{z}}$    & 1          & \tz              & \tz              & \tz              & \tz              & 1        \\
		$\mathbf{z}^{2}$  & 1          & \tz              & \tz              & \tz              & \tz              & 1        \\
		$\mathbf{z}^{3}$  & 2          & \tz              & \tz              & \tz              & \tz              & 2        \\
		$\mathbf{z}^{4}$  & 3          & 2                & \tz              & \tz              & \tz              & 5        \\
		$\mathbf{z}^{5}$  & 5          & 6                & \tz              & \tz              & \tz              & 11       \\
		$\mathbf{z}^{6}$  & 8          & 16               & 4                & \tz              & \tz              & 28       \\
		$\mathbf{z}^{7}$  & 14         & 40               & 20               & \tz              & \tz              & 74       \\
		$\mathbf{z}^{8}$  & 23         & 92               & 76               & 8                & \tz              & 199      \\
		$\mathbf{z}^{9}$  & 39         & 204              & 250              & 58               & \tz              & 551      \\
		$\mathbf{z}^{10}$ & 65         & 440              & 732              & 300              & 16               & 1553     \\
		$\mathbf{z}^{11}$ & 110        & 922              & 2000             & 1240             & 164              & 4436     \\
	\end{tabular}
	\caption{
		The enumeration of the class $\caut$.
		The sum column is equivalent to the enumeration of $\cat$.
	}
	\label{tbl:caut}
\end{table}

\subsubsection{Alkanes and Olefins}
We again turn our attention to the class containing both alkanes and non-allene olefins based on construction \ref{eq:ce}.
These molecules contain two kinds of stereogenic units: tetrahedral stereogenic centres, which we count with the variable $\uatom$, and stereogenic double bonds, which we count with $\vatom$.
This construction follows previous principles; the added complexity is limited due to the fact that a carbon cannot be involved in both type of stereogenic unit.
We start by considering the class counting all stereoisomers, as in Eq. \ref{eq:cetc}.
\begin{align}
	\ceutcFormula
	\label{eq:ceutc}
\end{align}
The generating function of this class is a generating function in three variables, which does not display naturally in a matrix.
Instead, we can count only the tetrahedral stereogenic centres or only the stereogenic double bonds separately, by setting $\vatom \gets 1$ or $\uatom \gets 1$, respectively, resulting in a bivariate generating function.
Alternatively, we can obtain a bivariate generating function counting the total number of stereogenic units with $\watom$ by setting $\vatom \gets \watom, \uatom \gets \watom$.
The first few terms are presented in tabular form in Table {\ref{tbl:ceutc}}.
An enumeration up to size 20, as well as results taken up to either or both forms of stereoisomerism can be found in Appendix \ref{sec:appendix_enum}.
\begin{table}[h!]
	\footnotesize
	\parbox{.31\linewidth}{
		\begin{tabular}{l|rrrr}
			\,               & $\mathbf1$ & $\mathbf{u}^{1}$ & $\mathbf{u}^{2}$ & $\mathbf{u}^{3}$ \\ \hline
			$\mathbf{{z}}$   & 1          & \tz              & \tz              & \tz              \\
			$\mathbf{z}^{2}$ & 2          & \tz              & \tz              & \tz              \\
			$\mathbf{z}^{3}$ & 6          & \tz              & \tz              & \tz              \\
			$\mathbf{z}^{4}$ & 16         & 4                & \tz              & \tz              \\
			$\mathbf{z}^{5}$ & 51         & 18               & \tz              & \tz              \\
			$\mathbf{z}^{6}$ & 161        & 88               & 8                & \tz              \\
			$\mathbf{z}^{7}$ & 538        & 390              & 60               & \tz              \\
			$\mathbf{z}^{8}$ & 1824       & 1664             & 396              & 16               \\
		\end{tabular}
	}
	\parbox{.31\linewidth}{
		\begin{tabular}{l|rrrr}
			\,               & $\mathbf1$ & $\mathbf{v}^{1}$ & $\mathbf{v}^{2}$ & $\mathbf{v}^{3}$ \\ \hline
			$\mathbf{{z}}$   & 1          & \tz              & \tz              & \tz              \\
			$\mathbf{z}^{2}$ & 2          & \tz              & \tz              & \tz              \\
			$\mathbf{z}^{3}$ & 4          & 2                & \tz              & \tz              \\
			$\mathbf{z}^{4}$ & 12         & 8                & \tz              & \tz              \\
			$\mathbf{z}^{5}$ & 31         & 34               & 4                & \tz              \\
			$\mathbf{z}^{6}$ & 95         & 130              & 32               & \tz              \\
			$\mathbf{z}^{7}$ & 290        & 498              & 192              & 8                \\
			$\mathbf{z}^{8}$ & 926        & 1874             & 994              & 106              \\
		\end{tabular}
	}
	\parbox{.31\linewidth}{
		\begin{tabular}{l|rrrr}
			\,               & $\mathbf1$ & $\mathbf{w}^{1}$ & $\mathbf{w}^{2}$ & $\mathbf{w}^{3}$ \\ \hline
			$\mathbf{{z}}$   & 1          & \tz              & \tz              & \tz              \\
			$\mathbf{z}^{2}$ & 2          & \tz              & \tz              & \tz              \\
			$\mathbf{z}^{3}$ & 4          & 2                & \tz              & \tz              \\
			$\mathbf{z}^{4}$ & 8          & 12               & \tz              & \tz              \\
			$\mathbf{z}^{5}$ & 17         & 44               & 8                & \tz              \\
			$\mathbf{z}^{6}$ & 35         & 146              & 76               & \tz              \\
			$\mathbf{z}^{7}$ & 74         & 446              & 434              & 34               \\
			$\mathbf{z}^{8}$ & 154        & 1286             & 1982             & 478              \\
		\end{tabular}
	}
	\vspace{1mm}
	\caption{
		Matrix representation of the class $\ceutc$, showing the expansion with the second variable counting tetrahedral stereogenic centres ($\uatom$), stereogenic double bonds ($\vatom$), and both ($\watom$).
	}
	\label{tbl:ceutc}
\end{table}

\subsection{Expected Density of Stereogenic Units} \label{sec:percent_stereo}
Given the matrix representation of the truncated generating function, presented in the last section, we can calculate the expected density of stereogenic units as a function of the size of the molecule.
For tetrahedral stereogenic centres, we denote this by $\frac{|a|_u}{|a|_z}$.
Given a class $\class{A}$, for molecules of size $n$, the expected value of this fraction is given by~\cite{blue_book}:

$$
	\mathbb{E} \left[ \frac{|a|_u}{|a|_z} \, \middle| \, |a|_z = n \right] = \frac{
	\left. [z^n] \frac{\partial}{\partial u} A(z, u) \right|_{u=1}
	}{
	n \, [z^n] A(z, 1)
	}
	.
$$
Here, taking the derivative with respect to $u$ multiplies each term by the exponent of $u$, thereby weighting each structure by its number of tetrahedral stereogenic centres.
Setting $u = 1$ then removes the marking while preserving the weights, and the numerator therefore gives the total number of tetrahedral stereogenic centres of the structures in $\class{A}_n$.
This is divided by the total size of the structures of $\class{A}_n$, obtained by extracting the number of structures from the corresponding univariate generating function (setting the parameter $u$ to $1$ in the generating function), and multiplying by the size.

\begin{figure}[H]
	\centering
	\begin{tikzpicture}
		\begin{axis}[PlotAxisStyle, ylabel={
		    $\uatommark / \atom$: tetrahedral stereogenic centres
		}, width=0.52\textwidth]
			\addplot[PlotStyle, color=black] table{data/enum_a_c.dat}; \label{plot:ca}
			\addplot[PlotStyle, color=colgreen,] table{data/enum_a_n.dat};\label{plot:cat}
		\end{axis}
		\legendtable{
			\ref{plot:ca}   & \cau         & (Eq. \ref{eq:cau}) \\
			\ref{plot:cat}  & \caut        & (Eq. \ref{eq:caut}) \\
		}
	\end{tikzpicture}
	\caption{
		For sizes up to $\atom = 20$, gives the fraction of carbon atoms which are tetrahedral stereogenic centres in the class of alkanes, up to stereochemistry (\cau), and counting stereoisomers (\caut).
	}
	\label{fig:countchiral_alkanes}
\end{figure}
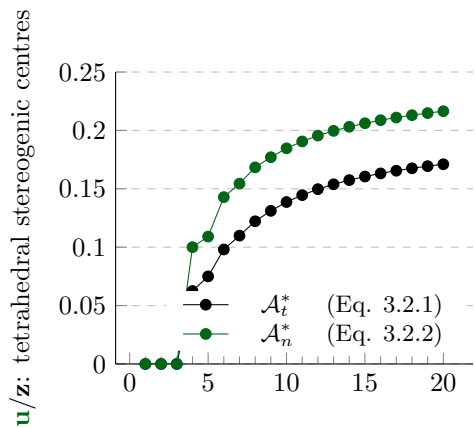

We begin with the alkanes.
For the classes $\cau$ and $\caut$, the results are presented in Fig. \ref{fig:countchiral_alkanes}.

% Graph - tetrahedral stereogenic centres in alkanes

Interestingly, the density of tetrahedral stereogenic centres differs significantly depending on whether the class of alkanes is taken up to stereoisomerism or not.
This can be explained by the following effect:
when constructing a molecule from the root, the presence of a stereogenic unit admits twice as many options for constructing subtrees, resulting in otherwise symmetrical structures with a greater presence of tetrahedral stereogenic centres being relatively over-represented when not taken up to symmetry.
\\

\begin{figure}[h]
	% Count u
	\begin{minipage}{0.49\textwidth}
		\begin{tikzpicture}
			\begin{axis}[PlotAxisStyle, ylabel={
			    a: $\uatommark / \atom$: tetrahedral stereogenic centres
			}, width=\textwidth]
				\addplot[PlotStyle, color=black]    table[x index=0, y index=1]{data/enum_e_gc.dat};     \label{plot:e_u_count_u}
				\addplot[PlotStyle, color=colgreen]     table[x index=0, y index=1]{data/enum_e_g.dat};   \label{plot:e_u_t_count_u}
				\addplot[PlotStyle, color=colred]      table[x index=0, y index=1]{data/enum_e_c.dat};   \label{plot:e_u_c_count_u}
				\addplot[PlotStyle, color=colblue]   table[x index=0, y index=1]{data/enum_e_n.dat};  \label{plot:e_u_tc_count_u}
			\end{axis}
		\end{tikzpicture}
	\end{minipage}
	% Count v
	\begin{minipage}{0.49\textwidth}
		\begin{tikzpicture}
			\begin{axis}[PlotAxisStyle, ylabel={
			    b: $\vatommark / \atom$: stereogenic double bonds
			}, width=\textwidth]
				\addplot[PlotStyle, color=black]    table[x index=0, y index=2]{data/enum_e_gc.dat};     \label{plot:e_u_count_v}
				\addplot[PlotStyle, color=colgreen]     table[x index=0, y index=2]{data/enum_e_g.dat};   \label{plot:e_u_t_count_v}
				\addplot[PlotStyle, color=colred]      table[x index=0, y index=2]{data/enum_e_c.dat};   \label{plot:e_u_c_count_v}
				\addplot[PlotStyle, color=colblue]   table[x index=0, y index=2]{data/enum_e_n.dat};  \label{plot:e_u_tc_count_v}
			\end{axis}
		\end{tikzpicture}
	\end{minipage}
	% Count UV
	\begin{minipage}{0.49\textwidth}
		\begin{tikzpicture}
			\begin{axis}[PlotAxisStyle, ylabel={
			    c: $\watomark / \atom$: stereogenic units
			}, width=\textwidth]
				\addplot[PlotStyle, color=black]    table[x index=0, y index=3]{data/enum_e_gc.dat};     \label{plot:e_u_count_uv}
				\addplot[PlotStyle, color=colgreen]     table[x index=0, y index=3]{data/enum_e_g.dat};   \label{plot:e_u_t_count_uv}
				\addplot[PlotStyle, color=colred]      table[x index=0, y index=3]{data/enum_e_c.dat};   \label{plot:e_u_c_count_uv}
				\addplot[PlotStyle, color=colblue]   table[x index=0, y index=3]{data/enum_e_n.dat};  \label{plot:e_u_tc_count_uv}
			\end{axis}
		\end{tikzpicture}
	\end{minipage}
	% Legend table
	\begin{minipage}{0.49\textwidth}
		\hspace{2mm} % Spacing from subfigure to the left
		\begin{tabular}{clr}
			\ref{plot:e_u_count_uv}    & \ceu   &                      \\ % (Eq. \ref{eq:ceu})
			\ref{plot:e_u_t_count_uv}  & \ceut  &                      \\ % (Eq. \ref{eq:ceut})
			\ref{plot:e_u_c_count_uv}  & \ceuc  &                      \\ % (Eq. \ref{eq:ceuc})
			\ref{plot:e_u_tc_count_uv} & \ceutc & (Eq. \ref{eq:ceutc}) \\
		\end{tabular}
	\end{minipage}
	\caption{
		For sizes up to $\atom = 20$, gives the expected density of tetrahedral stereogenic centres ($\uatom / \atom$), stereogenic double bonds ($\vatom / \atom$), and all stereogenic units ($\watom / \atom$) in the class of alkanes and olefins, up to tetrahedral and E/Z stereoisomerism (\ceu), tetrahedral stereoisomerism (\ceuc), E/Z stereoisomerism (\ceut), or neither (\ceutc).
	}
	\label{fig:countstereo}
\end{figure}
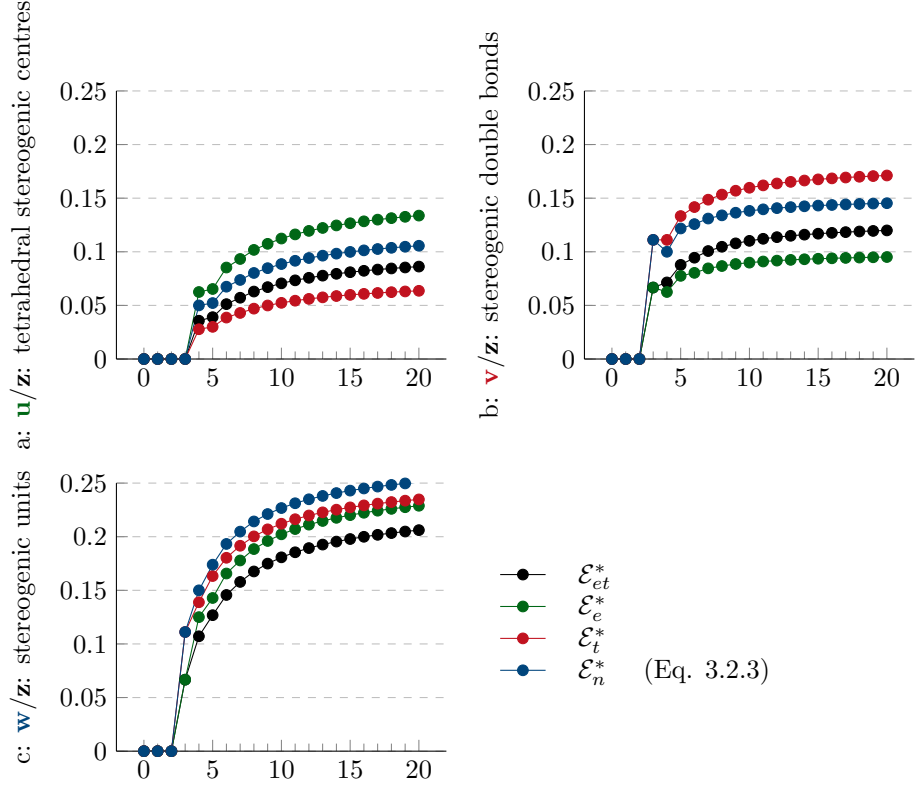

The impact of this effect becomes more complicated when considering the class of alkanes and olefins (Fig. \ref{fig:countstereo}), in which both tetrahedral and E/Z stereoisomerism need to be accounted for.
In Fig. \ref{fig:countstereo}a, in which tetrahedral stereogenic centres are counted, the effect is still clear when we compare the classes up to any isomerism (\ceu) and up to E/Z isomerism (\ceut).
However, comparing the class up to general stereoisomerism (\ceu) to the class up to tetrahedral stereoisomerism (\ceuc), we observe the inverse effect:
the presence of a stereogenic double bond does not cause the previously mentioned effect to count a greater presence of tetrahedral stereogenic centres, however the presence of such a bond does cause the involved atoms to be excluded from possibly being counted as tetrahedral stereogenic centres.
For this reason, fewer tetrahedral stereogenic centres are counted when counting up to E/Z but not tetrahedral stereoisomerism.
The same effect can be seen in Fig. \ref{fig:countstereo}b with the roles of tetrahedral stereogenic centres and stereogenic double bonds switched.

% Graph - in alkanes and olefins

Fig.~\ref{fig:countstereo}c shows the expected number of stereogenic units per carbon atom.
In this case, we observe a higher density when the class is taken up to neither form of stereoisomerism than both.
From this we can conclude that overcounting has a greater effect than the aforementioned inverse effect.
This is due to the fact that the inverse effect only confers an indirect `opportunity cost' on the number of stereogenic units, while the overcounting itself directly results in higher counts.
These results demonstrate that when stating general results about properties of combinatorial classes, one needs to be cognizant of possible symmetries of the structures and possible statistical biases which stem from these symmetries.

\subsection{Counting Achiral and Meso Molecules} \label{sec:achiral}
Finally, we can extend our approach to counting achiral molecules.
We define achiral molecules as those which are superposable on their mirror images.
Counting this class is not as simple as counting molecules with no tetrahedral stereogenic centres (such as taking the vertical expansion at $u = 0$ in previous classes), as we also need to count \emph{meso compounds}, achiral molecules containing chiral centres.
Fig.~\ref{fig:chiral_depends_on_subtree} provides an example of meso compounds (the two rightmost molecules):
in this case, the `root' carbon is a \emph{pseudoasymmetric centre}~\cite{nomenclature} - an atom which has two ligands which are enantiomers.

We first specify the class of achiral alkanes, $\caacht$, with the superscript $a$ denoting achirality, building upon specification \ref{eq:cat}.
\begin{align}
	\caachtFormula
	\label{eq:caacht}
\end{align}
In this specification, the class $\caacht$ defines the central, achiral `spine' of the compound, which can be thought of as a mirror plane.
For this reason, each $\caacht$ may produce at most one $\caacht$, ensuring linearity.
The $\diag_2(\cat)$ term denotes a pair of identical ligands attached to the `spine', with the specification of $\cat$ being identical to the previous specification; these are not necessarily achiral themselves, but up to combinatorial isomorphism, we may interpret these pairs as mirror image copies.

Enumerating on this specification gives the following sequence.
\countseq{\caacht}{
	0, 1, 1, 2, 3, 5, 8, 14, 23, 41, 69, 122, 208, 370, 636, 1134, 1963, 3505, 6099, 10908, 19059}

This is the OEIS sequence \oeis{A005627} (`Number of achiral planted trees with n nodes'), which was previously enumerated by R. W. Robinson, F. Harary, and A. T. Balaban in 1976~\cite{Robinson_Harary_Balaban_1976} by means of a system of recurrence relations.

The enumeration of achiral molecules includes meso compounds, but we also have the tools to explicitly count meso compounds.
With the superscript `$m$' denoting meso compounds, the enumeration is given by
\begin{align*}
	\camesot = \caacht \setminus \left( \left. \caut \right|_{u=0} \right)
	,
\end{align*}
which, within the class of molecules considered here, is exactly the achiral molecules ($\caacht$), excluding the molecules with no tetrahedral stereogenic centres ($ \left. \caut \right|_{u=0} $, specification \ref{eq:caut}).
This produces the following enumeration:
\countseq{\camesot}{
	0, 0, 0, 0, 0, 0, 0, 0, 0, 2, 4, 12, 24, 60, 116, 258, 492, 1030, 1940, 3912}
We observe that the first meso compounds occur at a size of $z = 9$; these are the two rightmost compounds illustrated in Fig~\ref{fig:chiral_depends_on_subtree}.
Additional specifications and enumerations for the number of achiral molecules for both alkanes as well as alkanes and olefins, taken up to the various kinds of isomerism are given in Appendix~\ref{sec:appendix_spec}.
For meso compounds, only alkanes are specified and enumerated, as identifying meso olefins would require comparing complete diastereomeric families, and is beyond the scope of the present work.

\section{Conclusion}
\label{sec:conc}
In this work, we have shown that analytic combinatorics offers a powerful, unified framework for enumeration problems, and applied this to the enumeration of various classes of chemical structures, reproducing classical results (sections 3.1.1, 3.4), and extending beyond these (sections 3.1.2, 3.2, 3.3, 3.4).
By expressing classes of chemical structures using combinatorial specifications, we use our tool \toolname in order to automatically obtain generating functions via the symbolic transfer theorems, and produce univariate and multivariate counting sequences.

We believe that \toolname can provide a platform for future work involving a wide variety of counting problems, not only in chemistry.
\toolname is being continuously developed, and we plan on extending the implementation to cover an even broader subset of the analytic combinatorics framework, for instance by supporting labelled structures and extending the set of allowed constructions to include the substitution and pointing operators.
The latter could allow for a similar approach to count acyclic molecules for which the tree structure does not have a defined `root'.
Taken more broadly, this study highlights the potential of analytic combinatorics to simplify a variety of combinatorial problems, and in combinatorial chemistry in particular.

\section*{Statements and Declarations}
\label{sec:decl}

\subsection*{Acknowledgements}
We thank Jonas Dørmann Vistrup for his contribution to the initial proof-of-concept version of \toolname.
We would additionally like to thank Natasja Find Jørgensen whose master's thesis, specifying some of these combinatorial classes, served as an inspiration for this work.

\subsection*{Code and Data Availability}
A Python notebook is available at \href{https://gitlab.com/casbjorn/molenum}{gitlab.com/casbjorn/molenum} to reproduce the results given in this paper, or as a demonstration of \toolname.
\toolname is available at \href{https://gitlab.com/casbjorn/combol}{gitlab.com/casbjorn/combol}.

\subsection*{Competing Interests}
The authors declare no conflict of interest.

\subsection*{Author Contributions}
Casper Asbjørn Eriksen:
Conceptualization,
methodology (lead),
formal analysis,
software,
validation,
writing - original draft%
.
Daniel Merkle:
Conceptualization,
methodology (supporting),
validation,
supervision,
writing - review \& editing%
.

\subsection*{Funding}
This work was supported by the Novo Nordisk Foundation (Grant numbers NNF19OC0057834 and NNF21OC0066551).

% Bibliography
\newpage
\bibliographystyle{plain}
\bibliography{refs}

% Appendices
\newpage
\begin{appendices}

	% Class specifications

\section{Class Specifications}
\label{sec:appendix_spec}
This appendix contains the specifications of all classes mentioned in this paper.
\subsection{Enumerating molecules}

\paragraph*{\ca: Alkanes, up to Tetrahedral Stereoisomerism}
\begin{align*}
	\caFormula
\end{align*}

\paragraph*{\cat: Alkanes}
\begin{align*}
	\catFormula
\end{align*}

\paragraph*{\ce: Alkanes and olefins, up to E/Z and Tetrahedral Stereoisomerism}
\begin{align*}
	\ceFormula
\end{align*}

\paragraph*{\cet: Alkanes and Olefins, up to E/Z Stereoisomerism}
\begin{align*}
	\cetFormula
\end{align*}

\paragraph*{\cec: Alkanes and Olefins, up to Tetrahedral Stereoisomerism}
\begin{align*}
	\cecFormula
\end{align*}

\paragraph*{\cetc: Alkanes and Olefins}
\begin{align*}
	\cetcFormula
\end{align*}

\subsection{Counting Stereogenic Units}

\paragraph*{\cau: Alkanes, up to Tetrahedral Stereoisomerism}
\begin{align*}
	\cauFormula
\end{align*}

\paragraph*{\caut: Alkanes}
\begin{align*}
	\cautFormula
\end{align*}

\paragraph*{\ceu: Alkanes and Olefins, up to E/Z and Tetrahedral Stereoisomerism}
\begin{align*}
	\ceuFormula
\end{align*}

\paragraph*{\ceut Alkanes and Olefins, up to E/Z Stereoisomerism}
\begin{align*}
	\ceutFormula
\end{align*}

\paragraph*{\ceuc: Alkanes and Olefins, up to Tetrahedral Stereoisomerism}
\begin{align*}
	\ceucFormula
\end{align*}

\paragraph*{\ceutc: Alkanes and Olefins}
\begin{align*}
	\ceutcFormula
\end{align*}

\subsection{Enumerating Achiral Molecules}

\paragraph*{\caach: Alkanes, up to Tetrahedral Stereoisomerism}
\begin{align*}
	\caachFormula
\end{align*}

\paragraph*{\caacht: Alkanes}
\begin{align*}
	\caachtFormula
\end{align*}

\paragraph*{\ceach: Alkanes and Olefins, up to E/Z and Tetrahedral Stereoisomerism}
\begin{align*}
	\ceachFormula
\end{align*}

\paragraph*{\ceacht Alkanes and Olefins, up to E/Z Stereoisomerism}
\begin{align*}
	\ceachtFormula
\end{align*}

\paragraph*{\ceachc: Alkanes and Olefins, up to Tetrahedral Stereoisomerism}
\begin{align*}
	\ceachcFormula
\end{align*}

\paragraph*{\ceachtc: Alkanes and Olefins}
\begin{align*}
	\ceachtcFormula
\end{align*}

	\section{Enumerations}
\label{sec:appendix_enum}
This appendix contains tables of enumerations up to greater sizes than could reasonably fit in the main paper.
All data contained here were generated using the supplementary notebook (\href{https://gitlab.com/casbjorn/molenum}{gitlab.com/casbjorn/molenum}), and are also included in the supplementary materials.

\FloatBarrier
\subsection{Enumerating molecules}
\begin{table}[h!]
    \ca: Alkanes , up tetrahedral stereoisomerism.
	\cat: Alkanes.
    \centering
    \footnotesize
    \begin{tabular}{rrr}
        $z$ & \ca & \cat \\
        \hline
        \textbf{0} & 0 & 0 \\
        \textbf{1} & 1 & 1 \\
        \textbf{2} & 1 & 1 \\
        \textbf{3} & 2 & 2 \\
        \textbf{4} & 4 & 5 \\
        \textbf{5} & 8 & 11 \\
        \textbf{6} & 17 & 28 \\
        \textbf{7} & 39 & 74 \\
        \textbf{8} & 89 & 199 \\
        \textbf{9} & 211 & 551 \\
        \textbf{10} & 507 & 1553 \\
        \textbf{11} & 1238 & 4436 \\
        \textbf{12} & 3057 & 12832 \\
        \textbf{13} & 7639 & 37496 \\
        \textbf{14} & 19241 & 110500 \\
        \textbf{15} & 48865 & 328092 \\
        \textbf{16} & 124906 & 980491 \\
        \textbf{17} & 321198 & 2946889 \\
        \textbf{18} & 830219 & 8901891 \\
        \textbf{19} & 2156010 & 27012286 \\
        \textbf{20} & 5622109 & 82300275 \\
        \textbf{21} & 14715813 & 251670563 \\
        \textbf{22} & 38649152 & 772160922 \\
        \textbf{23} & 101821927 & 2376294040 \\
        \textbf{24} & 269010485 & 7333282754 \\
        \textbf{25} & 712566567 & 22688455980 \\
        \textbf{26} & 1891993344 & 70361242924 \\
        \textbf{27} & 5034704828 & 218679264772 \\
        \textbf{28} & 13425117806 & 681018679604 \\
        \textbf{29} & 35866550869 & 2124842137550 \\
        \textbf{30} & 95991365288 & 6641338630714 \\
        \textbf{31} & 257332864506 & 20792003301836 \\
        \textbf{32} & 690928354105 & 65193446172901 \\
        \textbf{33} & 1857821351559 & 204709353135917 \\
        \textbf{34} & 5002305607153 & 643665829838389 \\
        \textbf{35} & 13486440075669 & 2026461371823166 \\
        \textbf{36} & 36404382430278 & 6387637263287353 \\
        \textbf{37} & 98380779170283 & 20157546705808565 \\
        \textbf{38} & 266158552000477 & 63680191033811326 \\
        \textbf{39} & 720807976831447 & 201379876145388644 \\
        \textbf{40} & 1954002050661819 & 637456295966779429 \\
    \end{tabular}
\end{table}

\begin{table}[h!]
	\ce: Alkanes and olefins, up to E/Z and tetrahedral stereoisomerism.
	\cet: Alkanes and olefins, up to E/Z stereoisomerism.
	\cec: Alkanes and olefins, up to tetrahedral stereoisomerism.
	\cetc: Alkanes and olefins.
	\centering
	\footnotesize
	\begin{tabular}{rrrrr}
        $z$ & \ce & \cet & \cec & \cetc \\
        \hline
        \textbf{0} & 0 & 0 & 0 & 0 \\
        \textbf{1} & 1 & 1 & 1 & 1 \\
        \textbf{2} & 2 & 2 & 2 & 2 \\
        \textbf{3} & 5 & 5 & 6 & 6 \\
        \textbf{4} & 14 & 16 & 18 & 20 \\
        \textbf{5} & 41 & 49 & 60 & 69 \\
        \textbf{6} & 127 & 168 & 207 & 257 \\
        \textbf{7} & 410 & 588 & 748 & 988 \\
        \textbf{8} & 1353 & 2125 & 2757 & 3900 \\
        \textbf{9} & 4576 & 7840 & 10411 & 15738 \\
        \textbf{10} & 15732 & 29468 & 39927 & 64581 \\
        \textbf{11} & 54874 & 112258 & 155307 & 268672 \\
        \textbf{12} & 193639 & 432870 & 610911 & 1130693 \\
        \textbf{13} & 690206 & 1685569 & 2426597 & 4804808 \\
        \textbf{14} & 2480944 & 6619746 & 9718055 & 20587994 \\
        \textbf{15} & 8983629 & 26188372 & 39199292 & 88854104 \\
        \textbf{16} & 32738382 & 104270256 & 159106666 & 385898644 \\
        \textbf{17} & 119980438 & 417500484 & 649376668 & 1685306216 \\
        \textbf{18} & 441907696 & 1680069099 & 2663379559 & 7396489878 \\
        \textbf{19} & 1634900899 & 6791060384 & 10971716397 & 32605260910 \\
        \textbf{20} & 6072852892 & 27560881309 & 45376369228 & 144302156477 \\
        \textbf{21} & 22639586343 & 112260107394 & 188336950168 & 640938804587 \\
        \textbf{22} & 84678626063 & 458764895447 & 784243095363 & 2856122812943 \\
        \textbf{23} & 317675277518 & 1880449023530 & 3275312151315 & 12765278562164 \\
        \textbf{24} & 1195054404545 & 7729117184199 & 13716212001164 & 57209589064643 \\
        \textbf{25} & 4507028566536 & 31849209336260 & 57583863974866 & 257038080134767 \\
        \textbf{26} & 17037513370281 & 131547502179198 & 242308716839136 & 1157529846253063 \\
        \textbf{27} & 64544586785104 & 544510207091578 & 1021797150202453 & 5223942041055886 \\
        \textbf{28} & 245009966991197 & 2258402633843389 & 4317392141366529 & 23622685821020448 \\
        \textbf{29} & 931789078705641 & 9384435117840768 & 18275954658958438 & 107020069029754350 \\
        \textbf{30} & 3549826419586126 & 39063567329058774 & 77497369090839167 & 485682493762153560 \\
        \textbf{31} & 13545755931403148 & 162870919222061989 & 329150015588918518 & 2207711884427361812 \\
        \textbf{32} & 51768027070667080 & 680109104951627832 & 1400093641246420172 & 10050578852399537506 \\
        \textbf{33} & 198126191135346317 & 2844045933673001083 & 5963988480618753480 & 45820360867369101648 \\
        \textbf{34} & 759289376585150875 & 11909150025075049789 & 25438807878697070022 & 209174257130569090562 \\
        \textbf{35} & 2913562421188542230 & 49931821527684421739 & 108643514830347619230 & 956107981984887450424 \\
        \textbf{36} & 11193402667640585072 & 209602132031557566155 & 464544755035946982019 & 4375463750863120177906 \\
        \textbf{37} & 43051913344895897434 & 880860829615248451369 & 1988566365262092770557 & 20046179423102945861206 \\
        \textbf{38} & 165763646778274346927 & 3705835487034218406398 & 8521492594413375574064 & 91939824400958718946616 \\
        \textbf{39} & 638893718143224871724 & 15606594418806178087219 & 36553634262865364895735 & 422101026306892155344870 \\
        \textbf{40} & 2464837130462256140738 & 65788664621591895465392 & 156950535158661089896618 & 1939758959471481891108231 \\
    \end{tabular}
\end{table}

\FloatBarrier
\subsection{Counting Stereogenic Units}
For these tables, counting tetrahedral stereogenic centres, stereogenic double bonds, and both, refer to the supplementary materials.

\FloatBarrier
\subsection{Counting Achiral Molecules}

\begin{table}[h!]
    \centering
    \footnotesize
    \begin{tabular}{rrr}
        $z$ & $\mathcal{A}_t^a$ & $\mathcal{A}_n^a$ \\
        \hline
        \textbf{0} & 0 & 0 \\
        \textbf{1} & 1 & 1 \\
        \textbf{2} & 1 & 1 \\
        \textbf{3} & 2 & 2 \\
        \textbf{4} & 3 & 3 \\
        \textbf{5} & 5 & 5 \\
        \textbf{6} & 8 & 8 \\
        \textbf{7} & 14 & 14 \\
        \textbf{8} & 23 & 23 \\
        \textbf{9} & 40 & 41 \\
        \textbf{10} & 67 & 69 \\
        \textbf{11} & 116 & 122 \\
        \textbf{12} & 196 & 208 \\
        \textbf{13} & 339 & 370 \\
        \textbf{14} & 576 & 636 \\
        \textbf{15} & 997 & 1134 \\
        \textbf{16} & 1701 & 1963 \\
        \textbf{17} & 2942 & 3505 \\
        \textbf{18} & 5034 & 6099 \\
        \textbf{19} & 8708 & 10908 \\
        \textbf{20} & 14931 & 19059 \\
        \textbf{21} & 25826 & 34129 \\
        \textbf{22} & 44354 & 59836 \\
        \textbf{23} & 76719 & 107256 \\
        \textbf{24} & 131922 & 188576 \\
        \textbf{25} & 228183 & 338322 \\
        \textbf{26} & 392757 & 596252 \\
        \textbf{27} & 679357 & 1070534 \\
        \textbf{28} & 1170254 & 1890548 \\
        \textbf{29} & 2024215 & 3396570 \\
        \textbf{30} & 3489121 & 6008908 \\
        \textbf{31} & 6035299 & 10801816 \\
        \textbf{32} & 10408429 & 19139155 \\
        \textbf{33} & 18004151 & 34422537 \\
        \textbf{34} & 31063270 & 61074583 \\
        \textbf{35} & 53732819 & 109894294 \\
        \textbf{36} & 92740844 & 195217253 \\
        \textbf{37} & 160423448 & 351404205 \\
        \textbf{38} & 276969008 & 624913284 \\
        \textbf{39} & 479106535 & 1125291874 \\
        \textbf{40} & 827383783 & 2003090071 \\
        \textbf{41} & 1431236971 & 3608175239 \\
        \textbf{42} & 2472187500 & 6428430129 \\
        \textbf{43} & 4276507618 & 11582995444 \\
        \textbf{44} & 7388228448 & 20653101216 \\
        \textbf{45} & 12780595559 & 37223637886 \\
        \textbf{46} & 22083705401 & 66420162952 \\
        \textbf{47} & 38201934221 & 119740546576 \\
        \textbf{48} & 66018618690 & 213802390264 \\
        \textbf{49} & 114204228437 & 385525375648 \\
        \textbf{50} & 197385592131 & 688796847976 \\
    \end{tabular}
\end{table}

\begin{table}[h!]
    \centering
    \footnotesize
    \begin{tabular}{rrrrr}
        $z$ & $\mathcal{E}_{et}^a$ & $\mathcal{E}_e^a$ & $\mathcal{E}_t^a$ & $\mathcal{E}_n^a$ \\
        \hline
        \textbf{0} & 0 & 0 & 0 & 0 \\
        \textbf{1} & 1 & 1 & 1 & 1 \\
        \textbf{2} & 2 & 2 & 2 & 2 \\
        \textbf{3} & 5 & 5 & 6 & 6 \\
        \textbf{4} & 12 & 12 & 16 & 16 \\
        \textbf{5} & 33 & 33 & 51 & 51 \\
        \textbf{6} & 90 & 90 & 161 & 161 \\
        \textbf{7} & 260 & 260 & 538 & 538 \\
        \textbf{8} & 759 & 759 & 1824 & 1824 \\
        \textbf{9} & 2278 & 2280 & 6340 & 6342 \\
        \textbf{10} & 6920 & 6924 & 22349 & 22353 \\
        \textbf{11} & 21362 & 21382 & 79929 & 79952 \\
        \textbf{12} & 66614 & 66668 & 288909 & 288979 \\
        \textbf{13} & 209912 & 210119 & 1054517 & 1054814 \\
        \textbf{14} & 666719 & 667339 & 3879770 & 3880804 \\
        \textbf{15} & 2133476 & 2135630 & 14376322 & 14380376 \\
        \textbf{16} & 6869267 & 6876120 & 53597656 & 53612702 \\
        \textbf{17} & 22242250 & 22265449 & 200915003 & 200973300 \\
        \textbf{18} & 72371582 & 72447819 & 756784066 & 757007106 \\
        \textbf{19} & 236526944 & 236784239 & 2862933260 & 2863801188 \\
        \textbf{20} & 776078878 & 776940198 & 10872745251 & 10876119999 \\
        \textbf{21} & 2555572536 & 2558490954 & 41437856257 & 41451078785 \\
        \textbf{22} & 8442667703 & 8452546852 & 158433081680 & 158484988549 \\
        \textbf{23} & 27974339023 & 28007984266 & 607526095335 & 607730701598 \\
        \textbf{24} & 92943469389 & 93058187782 & 2335870467507 & 2336678744539 \\
        \textbf{25} & 309572926687 & 309965430183 & 9003360102327 & 9006561470163 \\
        \textbf{26} & 1033493451119 & 1034838492291 & 34781661609204 & 34794365294837 \\
        \textbf{27} & 3457634246792 & 3462253819982 & 134652020058183 & 134702525224698 \\
        \textbf{28} & 11590718800246 & 11606608207830 & 522308413601342 & 522509513647806 \\
        \textbf{29} & 38926227366578 & 38980968583008 & 2029711380297943 & 2030513267714254 \\
        \textbf{30} & 130954774286731 & 131143604281676 & 7900998942867700 & 7904200504220362 \\
        \textbf{31} & 441264690879020 & 441916872550851 & 30805098155184183 & 30817895278042596 \\
        \textbf{32} & 1489128161138063 & 1491383069846608 & 120285404991219528 & 120336610284198738 \\
        \textbf{33} & 5032459561140142 & 5040263724098149 & 470342037967665535 & 470547122093961698 \\
        \textbf{34} & 17029723256576114 & 17056757465743871 & 1841575002071719431 & 1842397108652293618 \\
        \textbf{35} & 57700669565789900 & 57794396499283934 & 7219499058961983996 & 7222797231681302104 \\
        \textbf{36} & 195735557891147396 & 196060754973431887 & 28335883542165250284 & 28349125149987960028 \\
        \textbf{37} & 664731612301126930 & 665860724723494626 & 111339924790380781284 & 111393124162178953772 \\
        \textbf{38} & 2259877965168589013 & 2263800912634477471 & 437949622217242118670 & 438163492002255535284 \\
        \textbf{39} & 7690631325566393032 & 7704269406621853288 & 1724380099415672003301 & 1725240400664645277426 \\
        \textbf{40} & 26197233280583515586 & 26244672899161624181 & 6796037417484589458845 & 6799499941065329802351 \\
        \textbf{41} & 89319231546656380937 & 89484336595834894961 & 26808487726724923827900 & 26822430866013676702891 \\
        \textbf{42} & 304797768196211520009 & 305372673137989369781 & 105843448477367632311714 & 105899623082960496686496 \\
        \textbf{43} & 1040970200003534567448 & 1042972988545459510723 & 418228256278418266963982 & 418454678173300654154824 \\
        \textbf{44} & 3558018271318229348037 & 3564998432236582633623 & 1653881093744562171492918 & 1654794121813607632040573 \\
        \textbf{45} & 12170436553831382192541 & 12194774060702986517973 & 6545179909317597274498344 & 6548863119899593267371806 \\
        \textbf{46} & 41659877067602428117236 & 41744767154120180358432 & 25920963836927111534157024 & 25935827845721760982380236 \\
        \textbf{47} & 142702080809810610430745 & 142998290236250231545812 & 102725842593499541012152733 & 102785849798508249997557120 \\
        \textbf{48} & 489137102129691270387210 & 490171037231982812707964 & 407375464731751559729508043 & 407617802244178780004012669 \\
        \textbf{49} & 1677672308837434213500426 & 1681282517927340984375871 & 1616534421503186718305555757 & 1617513415626637793378228085 \\
        \textbf{50} & 5757697602510389967523101 & 5770307421553575905401068 & 6418580808612277024056445176 & 6422536976152525073994046048 \\
    \end{tabular}
\end{table}

\FloatBarrier
\subsection{Counting Meso Compounds}

\begin{table}[h!]
    \centering
    \footnotesize
    \begin{tabular}{rrr}
        $z$ & $\mathcal{A}_t^m$ & $\mathcal{A}_n^m$ \\
        \hline
        \textbf{0} & 0 & 0 \\
        \textbf{1} & 0 & 0 \\
        \textbf{2} & 0 & 0 \\
        \textbf{3} & 0 & 0 \\
        \textbf{4} & 0 & 0 \\
        \textbf{5} & 0 & 0 \\
        \textbf{6} & 0 & 0 \\
        \textbf{7} & 0 & 0 \\
        \textbf{8} & 0 & 0 \\
        \textbf{9} & 1 & 2 \\
        \textbf{10} & 2 & 4 \\
        \textbf{11} & 6 & 12 \\
        \textbf{12} & 12 & 24 \\
        \textbf{13} & 29 & 60 \\
        \textbf{14} & 56 & 116 \\
        \textbf{15} & 121 & 258 \\
        \textbf{16} & 230 & 492 \\
        \textbf{17} & 467 & 1030 \\
        \textbf{18} & 875 & 1940 \\
        \textbf{19} & 1712 & 3912 \\
        \textbf{20} & 3172 & 7300 \\
        \textbf{21} & 6051 & 14354 \\
        \textbf{22} & 11110 & 26592 \\
        \textbf{23} & 20817 & 51354 \\
        \textbf{24} & 37938 & 94592 \\
        \textbf{25} & 70153 & 180292 \\
        \textbf{26} & 127061 & 330556 \\
        \textbf{27} & 232611 & 623788 \\
        \textbf{28} & 419126 & 1139420 \\
        \textbf{29} & 761275 & 2133630 \\
        \textbf{30} & 1365677 & 3885464 \\
        \textbf{31} & 2464981 & 7231498 \\
        \textbf{32} & 4405446 & 13136172 \\
        \textbf{33} & 7910892 & 24329278 \\
        \textbf{34} & 14092839 & 44104152 \\
        \textbf{35} & 25199229 & 81360704 \\
        \textbf{36} & 44765463 & 147241872 \\
        \textbf{37} & 79759119 & 270739876 \\
        \textbf{38} & 141342724 & 489287000 \\
        \textbf{39} & 251068783 & 897254122 \\
        \textbf{40} & 443968660 & 1619674948 \\
        \textbf{41} & 786575062 & 2963513330 \\
        \textbf{42} & 1388274111 & 5344516740 \\
        \textbf{43} & 2454050590 & 9760538416 \\
        \textbf{44} & 4324008696 & 17588881464 \\
        \textbf{45} & 7628515615 & 32071557942 \\
        \textbf{46} & 13421165941 & 57757623492 \\
        \textbf{47} & 23637021155 & 105175633510 \\
        \textbf{48} & 41529647514 & 189313419088 \\
        \textbf{49} & 73029266823 & 344350414034 \\
        \textbf{50} & 128155349783 & 619566605628 \\
    \end{tabular}
\end{table}

\end{appendices}

\end{document}